\documentclass[12pt]{article}%
\usepackage[nosort]{cite}
\usepackage{graphicx}
\usepackage{multicol}
\usepackage{amsfonts}
\usepackage{amssymb}
\usepackage{amsmath}
\usepackage{heck}
\usepackage{afterpage}
\usepackage{setspace}
\usepackage{verbatim}
\usepackage{color}
\usepackage{longtable}
\usepackage{float}
\usepackage{subcaption}
\usepackage{epsfig}
\usepackage{enumerate}
\usepackage{epstopdf}
\usepackage[enableskew, vcentermath]{youngtab}
\usepackage{adjustbox}
\usepackage{multirow}
\usepackage{tikz}
\usepackage[margin=1in]{geometry}
\usepackage{titletoc}
\usepackage{hyperref}
\usepackage{mathtools}%
\providecommand{\U}[1]{\protect\rule{.1in}{.1in}}
\pdfoutput=1
\newsavebox{\mysavebox}

\hypersetup{colorlinks,citecolor=black,filecolor=black,linkcolor=black,urlcolor=black}
\usetikzlibrary{decorations.markings}

\numberwithin{equation}{section}

\def\gmax{\mathfrak{g}_{\rm max}}

\usetikzlibrary{chains}
\allowdisplaybreaks
\tikzset{node distance=2em, ch/.style={circle,draw,on chain,inner sep=2pt},chj/.style={ch,join},every path/.style={shorten >=4pt,shorten <=4pt},line width=1pt,baseline=-1ex}

\newcommand{\ba}{\begin{eqnarray}}
\newcommand{\ea}{\end{eqnarray}}

\newcommand{\mf}{\mathfrak}

\newcommand{\be}{\begin{equation}}
\newcommand{\ee}{\end{equation}}
\tikzstyle{startstop} = [rectangle, rounded corners, minimum width=3cm, minimum height=1cm,text centered, draw=black, fill=blue!10]
\tikzstyle{startstop} = [rectangle, rounded corners, minimum width=3cm, minimum height=1cm,text centered, draw=black, fill=blue!10]
\tikzstyle{io} = [trapezium, trapezium left angle=70, trapezium right angle=110, minimum width=3cm, minimum height=1cm, text centered, draw=black, fill=blue!30]
\tikzstyle{process} = [rectangle, minimum width=3cm, minimum height=1cm, text centered, draw=black, fill=orange!30]
\tikzstyle{decision} = [diamond, minimum width=3cm, minimum height=1cm, text centered, draw=black, fill=green!30]
\tikzstyle{arrow} = [thick,->,>=stealth]
\tikzset{->-/.style={decoration={
  markings,
  mark=at position #1 with {\arrow[scale=2.4]{>}}},postaction={decorate}}}
\makeatletter \@addtoreset{equation}{section} \makeatother

\begin{document}

\date{July 2018}

\title{Fission, Fusion, and 6D RG Flows}

\institution{PENN}{\centerline{${}^{1}$Department of Physics and Astronomy, University of Pennsylvania, Philadelphia, PA 19104, USA}}

\institution{IAS}{\centerline{${}^{2}$School of Natural Sciences, Institute for Advanced Study, Princeton, NJ 08540, USA}}

\institution{BICOCCA}{\centerline{${}^{3}$Dipartimento di Fisica, Universit\`a di Milano-Bicocca, Milan, Italy}}

\institution{INFN}{\centerline{${}^{4}$INFN, sezione di Milano-Bicocca, Milan, Italy}}

\authors{Jonathan J. Heckman\worksat{\PENN}\footnote{e-mail: {\tt jheckman@sas.upenn.edu}},
Tom Rudelius\worksat{\IAS}\footnote{e-mail: {\tt rudelius@ias.edu}},
and Alessandro Tomasiello\worksat{\BICOCCA,\INFN}\footnote{e-mail: {\tt alessandro.tomasiello@unimib.it}}}

\abstract{We show that all known 6D SCFTs
can be obtained iteratively from an underlying set of UV progenitor theories through the processes
of ``fission'' and ``fusion.'' Fission consists of a tensor branch deformation followed by a special class of
Higgs branch deformations characterized by discrete and continuous homomorphisms into flavor symmetry algebras. Almost all 6D SCFTs
can be realized as fission products. The remainder can be constructed via one step of
fusion involving these fission products, whereby a single common flavor
symmetry of decoupled 6D SCFTs is gauged and paired with a new tensor multiplet
at the origin of moduli space, producing an RG flow ``in reverse'' to the UV.
This leads to a streamlined labeling scheme for all known 6D SCFTs
in terms of a few pieces of group theoretic data. The
partial ordering of continuous homomorphisms $\mathfrak{su}(2) \rightarrow \mathfrak{g}_{\text{flav}}$
for $\mathfrak{g}_{\text{flav}}$ a flavor symmetry
also points the way to a classification of 6D RG flows.}

\maketitle

\setcounter{tocdepth}{2}

\tableofcontents


\newpage

\section{Introduction \label{sec:INTRO}}

One of the central themes of quantum field theory (QFT)\ is the dynamics of a
physical system at short versus long distance scales. Starting
from a fixed point of the renormalization group (RG), it is often possible to
perturb the system, thereby reaching a new fixed point at long distances. An
outstanding open question is to understand the possible \textquotedblleft
UV\ progenitors\textquotedblright\ of a given IR\ fixed point. From this
perspective, it is natural to ask whether it is possible to determine the full
network of possible connections between such fixed points.

This is clearly an ambitious goal, and in many cases, the best one can hope to
do is provide a coarse partial ordering of conformal field theories (CFTs) by a few
numerical quantities, such as the Euler conformal anomaly in even dimensions
(see e.g.~\cite{Zamolodchikov:1986gt, Cardy:1988cwa, Komargodski:2011vj}).
Indeed, the sheer number of quantum field theories which are known is
enormous and even determining a full list of conformal fixed points remains a
major area of investigation.

Perhaps surprisingly, this issue is tractable for 6D
superconformal field theories (SCFTs). The reason is that to even construct
examples of such theories, a number of delicate conditions need to be
satisfied. Long thought not to exist, the first examples of such theories
generated via string theory appeared in references \cite{Witten:1995zh,
Strominger:1995ac} (for the SCFT interpretation of these constructions,
see \cite{Seiberg:1996qx}), and by now there is a systematic method to construct and study such
models via F-theory on singular elliptically fibered Calabi-Yau threefolds \cite{Heckman:2013pva, DelZotto:2014hpa, Heckman:2015bfa}.
This method of construction appears to encompass all
previously known methods for realizing 6D\ SCFTs,
and suggests that this geometric classification is likely
complete.\footnote{A potential caveat to this statement is that there might exist theories without a tensor branch of moduli space, whereas all known theories have such a branch (for some discussion of this possibility, see e.g. \cite{Shimizu:2017kzs, Chang:2017xmr}).
Additionally, one must also allow for \textquotedblleft frozen\textquotedblright\ F-theory
backgrounds (see e.g.~\cite{deBoer:2001px, Tachikawa:2015wka, Bhardwaj:2018jgp}).
This adds a small number of additional examples, but all can be understood as quotients of a geometric
phase of F-theory \cite{Bhardwaj:2015xxa}.}
For a review of a top down approach to the
construction of 6D\ SCFTs in F-theory, see reference \cite{Heckman:2018jxk}.

One of the main results from references \cite{Heckman:2013pva, Heckman:2015bfa}
is that there is a rather rigid structure for such F-theory realized 6D\ SCFTs. All known theories admit
a tensor branch. After deformation onto a partial tensor branch, i.e., by giving expectation values to
those tensor multiplet scalars which canonically pair with ADE gauge algebras (with F-theory
fiber types $I_n$, $I_n^{\ast}$ and $II^\ast$, $III^\ast$, $IV^\ast$), the
6D field theory resembles a generalization of a quiver, with a single spine of
gauge groups connected by generalizations of hypermultiplets known as
\textquotedblleft6D\ conformal matter\textquotedblright\ \cite{DelZotto:2014hpa, Heckman:2014qba}.
With this list of theories in place, more refined questions become accessible such as the
possible interconnections associated with deforming one fixed point to another.
This circle of ideas has been developed in references \cite{Heckman:2015ola, Heckman:2015axa,
Heckman:2016ssk, Cremonesi:2015bld, Mekareeya:2016yal}.

There are two basic ways to flow to a new fixed point in six dimensions whilst
preserving $\mathcal{N} = (1,0)$ supersymmetry
involving the geometric operations of complex structure deformations of the Calabi-Yau threefold
and K\"{a}hler deformations of the base of the elliptic threefold. A complex
structure deformation corresponds to motion on the Higgs branch, while
a K\"{a}hler deformation specifies a tensor branch deformation.
This is corroborated both in holography \cite{Louis:2015mka} and field theory \cite{Cordova:2016xhm},
which shows that the only supersymmetric flows between 6D SCFTs are via
operator vevs.

To a large extent, the F-theory approach to 6D\ SCFTs is especially well
suited to the study of tensor branch flows. This is because the classification
results of reference \cite{Heckman:2015bfa} explicitly list the structure of the tensor
branch, and the earlier reference \cite{Heckman:2013pva} classifies the resulting singular
geometries after blowdown of all compact curves in the base.

Higgs branch flows can be understood as deformations of the minimal
Weierstrass model, but explicitly characterizing admissible deformations of
the geometry is still a challenging task. Reference \cite{DelZotto:2014hpa} proposed that
many such deformations can be understood in algebraic terms, either as
nilpotent orbits in a semi-simple Lie algebra, or as homomorphisms
from finite subgroups of $SU(2)$ to the group $E_{8}$. One of the interesting
features of nilpotent orbits is that they automatically come with a
partial ordering, and indeed, this ordering matches up (contravariantly) with Higgs
branch flows \cite{Heckman:2016ssk, Mekareeya:2016yal}.

A priori, there could be many RG\ flow trajectories from a pair of UV /
IR\ theories. Using our geometric characterization of 6D\ SCFTs, we find that
if such flows exist, there is a trajectory in which one first moves on the
tensor branch, and only then moves on the Higgs branch. Of course, there may
be other trajectories to the same fixed point and these can involve an
alternating sequence of tensor and Higgs branch flows.

Given the uniform quiver-like structure for most 6D\ SCFTs on a
(partial)\ tensor branch, it is perhaps not altogether surprising that there
is an underlying set of common progenitor theories for nearly all 6D\ SCFTs.
In M-theory terms, these are the theory of $k$ small instantons probing a
$\mathbb{C}^{2}/\Gamma_{ADE}$ orbifold singularity filled by an $E_{8}$ nine-brane
wall. Here, $\Gamma_{ADE}\subset SU(2)$ is a finite subgroup, as classified by
the ADE\ series. In F-theory terms, these configurations are given by a
collection of $k$ collapsing curves wrapped by $\mathfrak{g}_{ADE}$
7-branes according to the configuration:%
\begin{equation}
\lbrack E_{8}],\underset{k}{\underbrace{\overset{\mathfrak{g}_{ADE}%
}{1},\overset{\mathfrak{g}_{ADE}}{2},...,\overset{\mathfrak{g}_{ADE}}{2}}%
},[G_{ADE}], \label{smallinstbase}%
\end{equation}
where here, we have a single self-intersection $-1$ curve and $(k-1)$
self-intersection $-2$ curves which intersect as indicated in the diagram. The bracketed
groups on the left and right indicate flavor symmetries of the SCFT.\footnote{There is
also an $SU(2)_{\rm L}$ flavor symmetry, which is more manifest in the heterotic picture.}
We call this the rank $k$ theory of $(E_{8},G_{ADE})$ orbi-instantons. The F-theory description
of these models was studied first in reference \cite{Aspinwall:1997ye}.

We find that nearly all 6D\ SCFTs with a quiver-like description can be described in two steps:

\begin{itemize}
\item Step 1:\ Either perform a tensor branch flow of an $(E_{8},G_{ADE})$
orbi-instanton theory, or keep the original tensor branch.

\item Step 2: Perform a ``homplex deformation'' namely a
Higgs branch flow associated with decorating the left and
right of the new theory with algebraic data such as a choice of nilpotent
orbit or discrete group homomorphism $\Gamma_{ADE} \rightarrow E_8$ with
$\Gamma_{ADE} \subset SU(2)$ a finite subgroup.
\end{itemize}

At the very least, this allows us to understand the vast
majority of 6D\ SCFTs as flows from a very simple underlying set of
progenitor theories. Because the underlying process of a tensor branch flow
often produces more than one decoupled SCFT\ in the IR, we refer to the
above process as \textquotedblleft fission.\textquotedblright\

But some theories do not arise as a fission product of $(E_{8},G_{ADE})$
orbi-instantons. Rather, they involve moving back to the UV via an operation
we call \textquotedblleft fusion.\textquotedblright\ This involves taking at least one 6D SCFT,
but possibly multiple decoupled 6D SCFTs, gauging a common flavor symmetry, and pairing the new
non-abelian vector multiplet with a tensor multiplet with scalar sent to the origin of moduli space.
Taking the full list of fusion products, we obtain theories already encountered (via fission from the
theory of $(E_{8},G_{ADE})$ orbi-instantons) as well as a new class of
UV\ progenitor theories.

Starting from such fusion products we can
in principle iterate further by additional tensor branch flows and Higgs
branch deformations. A priori, the combination of fission and fusion operations
could then lead to a wild proliferation in possible IR fixed points.

\begin{figure}[t!]
\begin{center}
\includegraphics[trim={0cm 2cm 0cm 0cm},clip,scale=0.5]{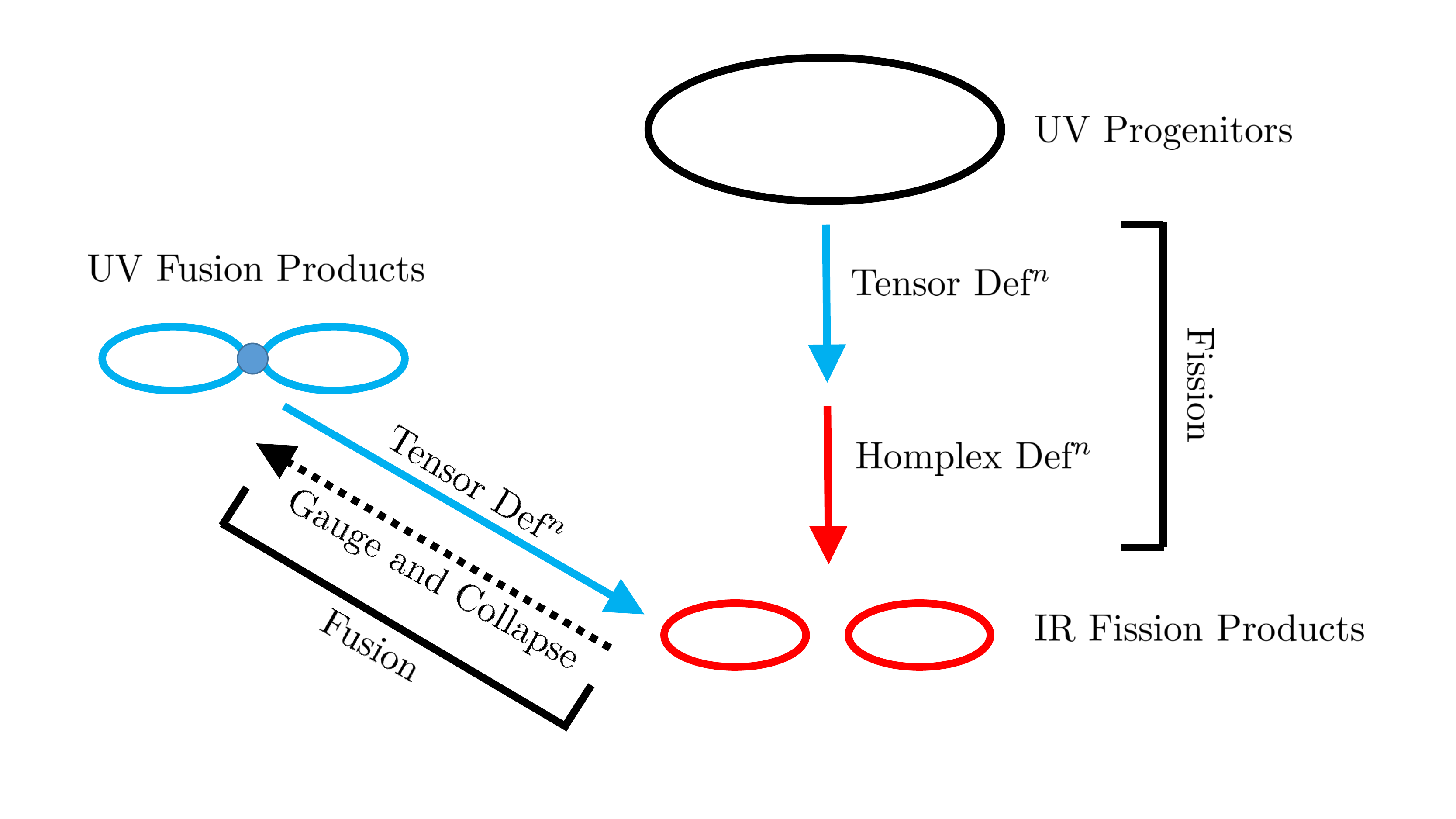}
\end{center}
\caption{Depiction of fission and fusion for 6D SCFTs. Fission consists of first performing a tensor branch deformation, which is
then followed by a specialized Higgs branch deformation associated with either a continuous $\mathfrak{su}(2) \rightarrow \mathfrak{g}_{\text{flav}}$ homomorphism or a homomorphism $\Gamma_{ADE} \rightarrow E_8$, with $\Gamma_{ADE}$ a finite subgroup of $SU(2)$.
In F-theory, these specify a restricted class of complex structure deformations which we refer to as ``homplex deformations.''
Fusion corresponds to gauging a flavor symmetry of at least one, but possibly several decoupled 6D SCFTs and pairing this gauge symmetry with a tensor multiplet. A new SCFT is generated by tuning this new tensor multiplet to the origin of moduli space. Quite surprisingly, all 6D SCFTs can be obtained from a single fission step or as the fusion of fission products obtained from a small number of UV progenitor theories.
Moreover, in a sense we will make precise below, almost all theories can be generated by fission alone.
}
\label{fig:FissionFusion}
\end{figure}

However, we find exactly the opposite. After precisely one fission step of an $(E_{8},G_{ADE})$
orbi-instanton and then possibly one fusion of such decay products, we obtain
all known 6D\ SCFTs. Already after one fission step one obtains almost
all theories, in a certain sense we will make precise below.
This yields a remarkably streamlined characterization of
6D SCFTs from a simple class of UV\ progenitor theories.

With these results in hand, we can also return to the original motivation for
this work: the classification of supersymmetric 6D RG flows. The results
we obtain provide a nearly complete characterization of ways to connect a
UV\ theory with candidate IR\ theories, though we do find some examples of
complex structure deformations which do not descend from a homplex deformation in the
sense of \textquotedblleft Step 2\textquotedblright\ outlined above.
Instead, these deformations spread across the entire generalized quiver, correlating
the flavor symmetry breaking pattern on the two sides. These deformations
are associated with semi-simple (that is, their matrix representatives are
diagonalizable) elements of the complexified flavor symmetry.
We find that for generic quiver-like theories, there is again a canonical ordering of such semi-simple elements,
as dictated by breaking patterns of gauge groups on the partial tensor branch.

The rest of this paper is organized as follows. First, in section \ref{sec:FISSFUSS} we
outline the general operations of fission and fusion for 6D SCFTs in F-theory.
Section \ref{sec:DARTHFISSIUS} shows that the vast majority of quiver-like
6D\ SCFTs are actually fission products of a small set
of UV progenitor theories. In section \ref{sec:DARTHFUSIUS} we
turn to theories generated by fusion, illustrating that in fact all 6D\ SCFTs
can be realized from at most one fission and fusion operation. Section
\ref{sec:FLOWS} discusses how the results of previous sections point
the way to a systematic treatment of 6D RG flows. We conclude in
section \ref{sec:CONC} and discuss some avenues for future investigation.

\section{Fission and Fusion for 6D\ SCFTs \label{sec:FISSFUSS}}

In this section we introduce two general operations for 6D\ SCFTs which we
refer to as fission and fusion. Fission corresponds to a tensor
branch deformation followed by a Higgs branch deformation characterized by discrete
or continuous homomorphisms. Fusion corresponds
to weakly gauging a common flavor symmetry of some
decoupled SCFTs, pairing the new vector multiplets
with a tensor multiplet (to cancel gauge anomalies)
and going to the origin of tensor branch moduli space.

To begin, let us briefly review some of the salient features of the geometric
realization of 6D\ SCFTs via backgrounds of F-theory. Following \cite{Heckman:2013pva},
we introduce an elliptically fibered Calabi-Yau threefold $X\rightarrow B$ in
which the base $B$ is non-compact. Recall that on a smooth base, we specify a
minimal Weierstrass model via:%
\begin{equation}
y^{2}=x^{3}+fx+g,
\end{equation}
with $f$ and $g$ sections of $K_{B}^{-4}$ and $K_{B}^{-6}$.

For the purposes of constructing 6D\ SCFTs, we seek configurations of
simultaneously contractible curves in the base $B$. A smooth base is constructed by joining the
non-Higgsable clusters of reference \cite{Morrison:2012np} with $-1$ curves
according to the specific SCFT gluing rules explained in reference \cite{Heckman:2013pva}.
In the limit where all curves in the configuration collapse to zero size, we obtain a 6D\ SCFT. The
classification results of reference \cite{Heckman:2013pva} determined that all such singular
limits for $B$ lead to orbifolds of the form $\mathbb{C}^{2}/\Gamma_{U(2)}$
for specific finite subgroups $\Gamma_{U(2)}\subset U(2)$.

The presentation of the Weierstrass model in this singular limit
is \cite{Morrison:2016nrt, DelZotto:2017pti}:
\begin{equation}
y^{2}=x^{3}+f_{\Gamma_{U(2)}}x+g_{\Gamma_{U(2)}},
\end{equation}
where the Weierstrass model parameters $f_{\Gamma_{U(2)}}$ and $g_{\Gamma_{U(2)}}$,
as well as $x$ and $y$ transform as $\Gamma_{U(2)}$ equivariant sections under the group action.
Explicitly, if we take elements $\gamma\in\Gamma_{U(2)}$ which act on local
$\mathbb{C}^{2}$ coordinates $(s,t)$ via $(s,t)\mapsto(\gamma_{s}%
(s,t),\gamma_{t}(s,t))$, the transformation rules for the $\Gamma_{U(2)}$ equivariant
sections of the Weierstrass model are:
\begin{align}
x  &  \mapsto\left(  \det\gamma\right)  ^{2}x\\
y  &  \mapsto\left(  \det\gamma\right)  ^{3}y\\
f(s,t)  &  \mapsto\left(  \det\gamma\right)  ^{4}f(\gamma_{s}(s,t),\gamma
_{t}(s,t))\\
g(s,t)  &  \mapsto\left(  \det\gamma\right)  ^{6}g(\gamma_{s}(s,t),\gamma
_{t}(s,t)).
\end{align}

In this presentation, non-abelian flavor symmetries are associated with
non-compact components of the discriminant locus; that is, they involve
7-branes wrapped on non-compact curves. Sometimes, a
6D\ SCFT\ may have a different flavor symmetry from what is indicated by the
Weierstrass model.

Reference \cite{Heckman:2015bfa} classified 6D SCFTs by determining all
possible F-theory backgrounds which can generate a 6D SCFT. This
was achieved by first listing all configurations of simultaneously contractible curves,
and then listing all possible elliptic fibrations over each
corresponding base.

The two presentations have their relative merits and provide complementary
perspectives on possible deformations of the geometry, which in turn
describe RG flows to new fixed points.

Starting from an F-theory model on a smooth base with a contractible configuration of curves,
we reach the singular limit described by an orbifold singularity by blowing down all curves of
self-intersection $-1$. Doing so shifts the self-intersection numbers of
curves which intersect such $-1$ curves which can in turn generate new curves
of self-intersection $-1$ after blowdown. Iterating in this way, one reaches
an \textquotedblleft endpoint configuration\textquotedblright\ in which no
$-1$ curves remain, that is to say, all remaining curves have
self-intersection $-m$ with $m>1$. The structure of these endpoint
configurations were completely classified in reference \cite{Heckman:2013pva}, and have
intersection pairings which are natural generalizations of the ADE\ series
associated with Kleinian singularities. In what follows, we shall often have
special need to reference the total number of curves in an endpoint
configuration, which we denote by $\ell_{\text{end}}$, in the obvious
notation. For example, in an A-type endpoint configuration we have:%
\begin{equation}
\underset{\ell_{\text{end}}}{\underbrace{m_{1},...,m_{\ell_{\text{end}}}}}.
\end{equation}

Our plan in the remainder of this section will be to characterize the
geometric content of RG\ flows for 6D\ SCFTs. We begin with a discussion of
tensor branch and Higgs branch deformations, and then turn to the specific
case of fission and fusion operations.

\subsection{Tensor Branch Deformations}

Tensor branch deformations correspond in the geometry to performing a blowup
of the base $B$. Recall that for any base $B$, we blowup at a point
$p$ of $B$ by introducing the space $B\times\mathbb{P}^{1}$ and defining a new
space Bl$_{p}B$ by the hypersurface:%
\begin{equation}
\widetilde{u}v_{B}+\widetilde{v}u_{B}=0,
\end{equation}
where $[\widetilde{u},\widetilde{v}]$ define homogeneous coordinates on the
$\mathbb{P}^{1}$ factor and $(u_{B},v_{B})$ are two sections of some bundle
defined on $B$ which have a zero at the point $p$. After the blowup
the canonical class of Bl$_{p}B$ is:
\begin{equation}
K_{\text{B}l_{p}B}=K_{B}+ E_{\text{new}},
\end{equation}
where $E_{\text{new}}$ denotes the class of our new exceptional divisor,
and the Weierstrass coefficients $f$ and $g$ are now sections of
$K_{\text{B}l_{p}B}^{-4}$ and $K_{\text{B}l_{p}B}^{-6}.$

We can consider more elaborate sequences of blowups by introducing additional
$\mathbb{P}^{1}$ factors, and in iterating in this way, we can view the new
base as the intersection of varieties in the ambient space:%
\begin{equation}
B^{(m)}=B\times\underset{m}{\underbrace{\mathbb{P}^{1}\times...\times
\mathbb{P}^{1}}},
\end{equation}
where $m$ indicates the number of blowups of the original base.

This presentation is especially helpful when we work in terms of a base given by an
orbifold $\mathbb{C}^{2}/\Gamma_{U(2)}$ and provides one way to implicitly
specify the new Weierstrass model after a tensor branch flow. Of course, we
can also work with all curves at finite size, and then we can simply indicate
which of these curves is to be decompactified at each stage. The disadvantage
of this description is that it does not provide us with an explicit
Weierstrass model for the global geometry, only one which is implicitly
specified patch by patch.

\subsection{Homplex Higgs Branch Deformations}

Higgs branch deformations are characterized by
perturbations $(\delta f , \delta g)$ in the Weierstrass model:%
\begin{equation}
y^{2}=x^{3}+(f+\delta f)x+(g+\delta g),
\end{equation}
such that the elliptic fibration becomes less singular after applying such a
perturbation. In the cases we shall consider in this paper, there is a close
interplay between breaking patterns of a flavor symmetry $\mathfrak{g}%
_{\text{flav}}$, and the associated unfolding of the singular fibration.
Recall that for generic unfoldings of a singularity, one specifies a Cartan
subalgebra of $\mathfrak{g}_{\text{flav}}$. Then, following the procedure in
\cite{Bershadsky:1996nh}, we can read off the unfolding of the singularity
(see also \cite{Katz:1996xe}).

Many Higgs branch deformations of 6D\ SCFTs can
be understood in terms of algebraic data associated with
breaking patterns of the flavor symmetry, and we refer to this special class
of deformations as \textquotedblleft homplex\textquotedblright\ Higgs
deformations, since they reference homomorphisms into the flavor symmetry algebra.

There are two cases in particular which figure prominently in the study of RG\ flows for 6D\ SCFTs:
\begin{itemize}
	\item Continuous homomorphisms $\mathfrak{su}(2)\rightarrow
	\mathfrak{g}_{\text{flav}}$.
	\item Group homomorphisms $\Gamma_{ADE}\rightarrow E_{8}$ with $\Gamma_{ADE}$ a finite subgroup of
	$SU(2)$.
\end{itemize}

Continuous homomorphisms $\mathfrak{su}(2)\rightarrow\mathfrak{g}%
_{\text{flav}}$ are all labeled by the orbits of nilpotent elements in
$\mathfrak{g}_{\text{flav}}$. Given a nilpotent element $\mu\in\mathfrak{g}%
_{\text{flav}}$, there is a corresponding $\mathfrak{su}(2)$ algebra, as
defined by $\mu$, $\mu^{\dag}$ and the commutator $[\mu,\mu^{\dag}]$. Even
though a nilpotent element defines a T-brane deformation\footnote{For a partial list of references
to T-branes in F-theory see e.g. \cite{Aspinwall:1998he, Donagi:2003hh, Cecotti:2010bp, Anderson:2013rka, Collinucci:2014taa, Collinucci:2014qfa, Bena:2016oqr, Marchesano:2016cqg, Anderson:2017rpr, Bena:2017jhm, Marchesano:2017kke, Cvetic:2018xaq}.}
of the SCFT \cite{DelZotto:2014hpa} (see also \cite{Heckman:2016ssk, Mekareeya:2016yal, Cvetic:2018xaq}),
the commutator $[\mu,\mu^{\dag}]$ is also a generator in the Cartan subalgebra, and can therefore be identified
with some unfolding of the singularity.\footnote{There are also many elements in the Cartan subalgebra which do not specify nilpotent elements, and in principle a full study of possible Higgs branch flows would need to also
address these different cases. In general, an element $\gamma$ of a semi-simple Lie algebra $\mathfrak{g}$ can be decomposed
into a semi-simple (meaning its matrix representations are diagonalizable) and nilpotent piece:
$\gamma = \gamma_{\text{semi}} + \gamma_{\text{nilp}}$. In the context of Higgs branch flows for 6D SCFTs, such semi-simple
deformations correlate deformations in a non-local way across a generalized quiver. We shall return
to some properties of such semi-simple flows in section \ref{sec:FLOWS}.} For classical
flavor symmetry algebras of $\mathfrak{su},\mathfrak{so},\mathfrak{sp}$-type,
we can label an orbit by a partition of integers $[\mu_1^{a_1},...,\mu_{k}^{a_k}]$,
where we take $\mu_1 > ... > \mu_k$, and $a_i > 0$ indicates the multiplicity of a given
integer. We shall also use the notation $\mu^{(m)}$ to denote a partition for the integer $m$,
and $\mu^T$ to denote the transpose of the Young diagram. For $\mathfrak{so},\mathfrak{sp}$ additional restrictions on admissible partitions apply. For the exceptional algebras, we instead use the Bala--Carter labels of a nilpotent orbit. See e.g.~\cite{NILPbook} as well as \cite{Chacaltana:2012zy} for additional details on nilpotent orbits.
	
Concretely, the theories associated to a given nilpotent Higgsing can be found by a variety of methods. For the $\mathfrak{su}$ case, a simple combinatorial method is available, summarized for example in \cite{Cremonesi:2015bld} and ultimately going back to \cite{Hanany:1996ie,Gaiotto:2008sa}. For the other Lie algebras, one can proceed for example by examining all the possible Higgs RG flows, and matching the resulting partial ordering to the partial ordering of nilpotent orbits via (Zariski)\ closure of orbits in the corresponding Lie algebra. We refer the reader to \cite{Heckman:2016ssk} for many examples of nilpotent Higgsing, including all the theories with $2\ldots 2$ endpoint; we will see some further examples in sections \ref{ssec:examples} and \ref{ssec:short copy(1)}.\footnote{Moreover, the dimensions of the orbits also match with anomaly and moduli space arguments \cite{Mekareeya:2016yal}; in the $\mathfrak{so}$ case, one can use this fact to provide a more direct combinatorial map between nilpotent elements and Higgsed theories.} So at least for this partial list of Higgs branch deformations, there is a classification of RG\ flows available.

The other algebraic data which prominently features in our analysis of Higgs
branch deformations comes from discrete group homomorphisms $\Gamma_{ADE}\rightarrow
E_{8}$ with $\Gamma_{ADE} \subset SU(2)$ a finite subgroup.
This shows up most naturally in the theory of heterotic small
instantons placed at an orbifold singularity $\mathbb{C}^{2}/\Gamma_{ADE}$, and
possible boundary conditions for the small instanton are specified by elements of
$\text{Hom}(\pi_{1}(S^3 / \Gamma_{ADE}), E_8) \simeq \text{Hom}( \Gamma_{ADE}, E_8)$,
each of which corresponds to an RG\ flow.
Many examples of theories obtained via this Higgsing were found in section 7 of reference
\cite{Heckman:2015bfa}, and an algorithm was proposed in \cite{Mekareeya:2017jgc, TRFrey}.
Again we will see some examples in section \ref{ssec:examples}.

There is no known purely mathematical partial ordering for such homomorphisms, but it is expected based
on physical considerations. At a crude level, we can see that at
least for those discrete group homomorphisms which define the same breaking pattern
as a continuous $\mathfrak{su}(2) \rightarrow \mathfrak{e}_8$ homomorphism, we
can simply borrow the partial ordering for nilpotent orbits. The subtlety in this approach is that
there are more discrete group homomorphisms than continuous $\mathfrak{su}(2) \rightarrow \mathfrak{e}_8$ homomorphisms.

Indeed, the appearance of these discrete homomorphisms is considerably more delicate
than their continuous group counterparts. For example, they only make an
appearance in the special case where a collapsing $-1$ curve enjoys an $E_{8}$
flavor symmetry (which may be emergent at the fixed point). In all
other cases where we have a curve of self-intersection $-n$ with $n>1$ which
enjoys an emergent flavor symmetry $\mathfrak{g}_{\text{flav}}$, the algebraic
data of a Higgs branch flow will always be associated with a continuous
homomorphism $\mathfrak{su}(2)\rightarrow\mathfrak{g}_{\text{flav}}$.

Putting this together, we see that instead of specifying all possible
deformations of the Weierstrass model, we can summarize many theories by algebraic data.
An additional benefit of this description is that the partial ordering for nilpotent orbits
coincides with that for 6D\ RG\ flows.

\subsection{Fission}

A general RG\ flow trajectory may consist of several steps of tensor branch
and Higgs branch deformations. The basic RG\ flow move we shall be interested
in for much of this paper is \textquotedblleft fission\textquotedblright%
\ which we define as a tensor branch deformation followed by a
homplex deformation in the sense defined in the previous section
(namely, by decoration by either discrete or continuous homomorphisms of the
flavor symmetry algebra). Note that we also allow the tensor branch and Higgs
branch deformation steps to be trivial, i.e., the case where we do no deformation at all.

The reason for the terminology is that after a tensor branch flow, we often get
decoupled 6D\ SCFTs. As an illustrative example, consider the F-theory
model with partial tensor branch:%
\begin{equation}
\lbrack E_{8}],\underset{k_{\mathrm{L}}}{\underbrace{\overset{\mathfrak{e}_{8}%
}{2},...,\overset{\mathfrak{e}_{8}}{2}}},\overset{\mathfrak{e}_{8}%
}{2},\underset{k_{\mathrm{R}}}{\underbrace{\overset{\mathfrak{e}_{8}}{2}%
,..,\overset{\mathfrak{e}_{8}}{2}}},[E_{8}], \label{ConfMattRankk}%
\end{equation}
which is also described by $k=k_{\mathrm{L}}+k_{\mathrm{R}}$ M5-branes probing an $E_{8}$
singularity. Decompactifying the middle $-2$ curve yields two decoupled
theories in the deep IR:%
\begin{gather}
\text{Example of Fission} \nonumber\\
\lbrack E_{8}],\underset{k_{\mathrm{L}}}{\underbrace{\overset{\mathfrak{e}_{8}%
}{2},...,\overset{\mathfrak{e}_{8}}{2}}},\overset{\mathfrak{e}_{8}%
}{2},\underset{k_{\mathrm{R}}}{\underbrace{\overset{\mathfrak{e}_{8}}{2}%
,..,\overset{\mathfrak{e}_{8}}{2}}},[E_{8}]\ \ \rightarrow\ \ \lbrack E_{8}%
],\underset{k_{\mathrm{L}}}{\underbrace{\overset{\mathfrak{e}_{8}}{2}%
,...,\overset{\mathfrak{e}_{8}}{2}}},[E_{8}]\oplus\lbrack E_{8}%
],\underset{k_{\mathrm{R}}}{\underbrace{\overset{\mathfrak{e}_{8}}{2}%
,..,\overset{\mathfrak{e}_{8}}{2}}},[E_{8}].  \label{FissionExample}
\end{gather}
We can then perform a further Higgs branch deformation associated with a
nilpotent orbit for each of our four $E_{8}$ factors (an analog of beta decay in nuclear physics).
Clearly, this is a \textquotedblleft fission operation.\textquotedblright\ We will show
in section \ref{sec:DARTHFISSIUS} that nearly all 6D\ SCFTs can be viewed as a fission product
from an $(E_{8},G_{ADE})$ orbi-instanton theory with partial tensor branch
description:%
\begin{equation}
\lbrack E_{8}],\underset{k}{\underbrace{\overset{\mathfrak{g}_{ADE}%
}{1},\overset{\mathfrak{g}_{ADE}}{2},...,\overset{\mathfrak{g}_{ADE}}{2}}%
},[G_{ADE}].
\end{equation}

In the case where there is no tensor branch flow, the algebraic Higgs branch
deformations will be labeled by a discrete homomorphism $\Gamma_{ADE}\rightarrow
E_{8}$ on the left (associated with the $-1$ curve touching the $E_{8}$
factor) and with a continuous homomorphism $\mathfrak{su}(2)\rightarrow
\mathfrak{g}_{ADE}$ on the right.

\subsection{Fusion}

We can also consider reversing the direction of an RG\ flow via a procedure we
refer to as a \textquotedblleft fusion operation.\textquotedblright\ We define
this as gauging a flavor symmetry for at least one, but possibly multiple decoupled SCFTs, and
introducing a single tensor multiplet in order to cancel the corresponding
gauge anomalies. Going to the origin of the tensor branch then takes us to a
new 6D\ SCFT which can clearly flow (after a tensor branch deformation)\ via a
fission operation back to the original set of decoupled SCFTs. As an example,
consider the reverse of the fission operation in the example of line
(\ref{FissionExample}):%
\begin{gather}
\text{Example of Fusion} \nonumber\\
\lbrack E_{8}],\underset{k_{\mathrm{L}}}{\underbrace{\overset{\mathfrak{e}_{8}%
}{2},...,\overset{\mathfrak{e}_{8}}{2}}},[E_{8}]\oplus\lbrack E_{8}%
],\underset{k_{\mathrm{R}}}{\underbrace{\overset{\mathfrak{e}_{8}}{2}%
,..,\overset{\mathfrak{e}_{8}}{2}}},[E_{8}]\ \ \rightarrow\ \ \lbrack E_{8}%
],\underset{k_{\mathrm{L}}}{\underbrace{\overset{\mathfrak{e}_{8}}{2}%
,...,\overset{\mathfrak{e}_{8}}{2}}},\overset{\mathfrak{e}_{8}}{2}%
,\underset{k_{\mathrm{R}}}{\underbrace{\overset{\mathfrak{e}_{8}}{2}%
,..,\overset{\mathfrak{e}_{8}}{2}}},[E_{8}].
\end{gather}

Having defined the basic operations of fission and fusion, we now
systematically study fission and fusion in 6D\ SCFTs.

\section{Nearly all 6D\ SCFTs as Fission Products \label{sec:DARTHFISSIUS}}

In this section we show that nearly all 6D\ SCFTs can be realized as fission
products of a handful of progenitor theories. We note that after excluding
models with a D-type endpoint, this includes all theories with a semi-classical
holographic dual. To accomplish this, we briefly
review some elements of the classification results in reference \cite{Heckman:2015bfa}.
We recall that the generic 6D\ SCFT can, on a partial tensor branch, be described
in terms of a generalized quiver-like theory. We will establish in this
section that these quiver-like theories all descend from the fission of a
simple class of progenitor theories labeled by the $(E_{8},G_{ADE})$
rank $k$ orbi-instanton theories.

This section is organized as follows. First, we briefly review the structural
elements of 6D\ SCFTs, particularly as quiver-like gauge theories. We then
introduce our progenitor theories and subsequently show that the products of fission
from these progenitors yields nearly all 6D SCFTs.

\subsection{The Quiver-like Structure of 6D\ SCFTs}

One of the main results of reference \cite{Heckman:2015bfa} is that the classification of
6D\ SCFTs which can be obtained from F-theory backgrounds can be split into
two steps. The first involves a classification of bases, and subsequently, we
can consider all possible ways of decorating a given base by singular elliptic
fibers. Quite remarkably, all bases resemble, on a partial
tensor branch, a quiver-like gauge theory. The main idea here is to split up
configurations of curves according to the algebra supported on a non-Higgsable
cluster. In particular, we have \textquotedblleft nodes\textquotedblright%
\ composed of the D / E-type algebras and corresponding self-intersection
number $-4$, $-6$, $-7$, $-8$, $-12$, with the remaining non-Higgsable
clusters, as well as the $-1$ curves used to build conformal matter
\textquotedblleft links\textquotedblright\ connecting the nodes. We can also
extend this classification terminology to include nodes where we have a $-2$
curve and a split $I_{m}$ fiber (i.e.~a 7-brane with $\mathfrak{su}(m)$
algebra) over a curve. In what follows, we can consider a quiver-like theory
to be one which admits a partial tensor branch with any of the ADE algebras
over these curves.

The resulting structure for all 6D\ SCFTs obtained in \cite{Heckman:2015bfa} is of the form
\begin{equation}
\lbrack G_{0}]-\overset{|}{G_{1}}-\overset{|}{G_{2}}-...-\underset{\ell
_{\text{plat}}}{\underbrace{G_{\text{max}}-...-G_{\text{max}}}}%
-...-\overset{|}{G_{\ell_{\text{quiv}}-1}}-\overset{|}{G_{\ell_{\text{quiv}}}%
}-[G_{\ell_{\text{quiv}}+1}], \label{generic}%
\end{equation}
where the $G_i$ are
ADE\ gauge group nodes, and the links \textquotedblleft$-$\textquotedblright\ are so-called
``conformal matter'' theories \cite{Heckman:2013pva, DelZotto:2014hpa, Heckman:2014qba}.\footnote{See also earlier work by \cite{Bershadsky:1996nu, Aspinwall:1997ye, Morrison:2012np}.} For example, the conformal matter theories connecting two copies of the same group
are given by:
\begin{subequations}\label{eq:cm}
\begin{align}
&  \mathfrak{e}_{8}\text{: \ \ }[E_{8}%
],1,2,\overset{\mathfrak{sp}_{1}}{2},\overset{\mathfrak{g}_{2}}{3}%
,1,\overset{\mathfrak{f}_{4}}{5},1,\overset{\mathfrak{g}_{2}}{3}%
,\overset{\mathfrak{sp}_{1}}{2},2,1,[E_{8}]\label{gmaxone}\\
&  \mathfrak{e}_{7}\text{: \ \ }[E_{7}%
],1,\overset{\mathfrak{su}_{2}}{2},\overset{\mathfrak{so}_{7}}{3}%
,\overset{\mathfrak{su}_{2}}{2},1,[E_{7}]\\
&\mathfrak{e}_{6}\text{: \ \ }[E_{6}%
],1,\overset{\mathfrak{su}_{3}}{3},1,[E_{6}]\\
&\mathfrak{so}_{2m}\text{: \ \ }[SO_{2m}%
],\overset{\mathfrak{sp}_{m-4}}{1},[SO_{2m}]\\
&\mathfrak{su}_{m}\text{: \ \ }[SU_{m}%
],[SU_{m}]. \label{gmaxfive}%
\end{align}
\end{subequations}

In line (\ref{generic}), generically (i.e.~for $\ell_{\text{quiv}}$ sufficiently large) only the two leftmost and
two rightmost nodes can attach to more than two links. We shall refer to
$\ell_{\text{quiv}}$ as the number of ADE gauge group nodes. An additional feature
is a nested sequence of containment relations for the
associated Lie algebras on each node. For some $i$ such that $1\leq
i_{\text{mid}}\leq\ell_{\text{quiv}}$, we have:%
\begin{equation}
\mathfrak{g}_{1}\subseteq...\subseteq\mathfrak{g}_{i_{\text{mid}}}%
\supseteq...\supseteq\mathfrak{g}_{\ell_{\text{quiv}}},
\end{equation}
so we can also assign the data $\mathfrak{g}_{\text{max}}$, a maximal gauge
algebra to each such theory. In many 6D\ SCFTs, this maximal algebra will
appear repeatedly on the \textquotedblleft plateau\textquotedblright\ of a
sequence of gauge algebras, and we label this quantity as $\ell_{\text{plat}}%
$. Because of the generic structure of such quiver-like theories, it will also
prove convenient to consider the \textquotedblleft analytic
continuation\textquotedblright\ of a given type of quiver to $\ell
_{\text{plat}}=0$ and even $\ell_{\text{plat}}=-1$. We can do
so when the structure of the ramps of gauge algebras on the left and right
admit such an extension. We stress that this is just a matter of notation, and we do not entertain a \textquotedblleft negative number of gauge
groups\textquotedblright\ as a physically meaningful notion.

Now, the structure of line (\ref{generic}) becomes most uniform when the number of gauge nodes
is sufficiently large. There are also 6D\ SCFTs which contain no gauge nodes
whatsoever, and are purely built from links (which were also classified in
reference \cite{Heckman:2015bfa}). These often do not fit into regular patterns of the kind
already introduced, but as we will shortly show, they can all instead be
viewed as the results of fusion operations.

With these elements in place, let us now turn to the progenitor theories which
produce, via fission, nearly all 6D\ SCFTs.

\subsection{Progenitor Theories}

We now introduce a small special class of progenitor theories from which we
construct nearly all 6D SCFTs as fission products. The theories in question
are the $(E_{8},G_{ADE})$ rank $k$ orbi-instanton theories with partial tensor
branch description:%
\begin{equation}
\lbrack E_{8}],\underset{k}{\underbrace{\overset{\mathfrak{g}_{ADE}%
}{1},\overset{\mathfrak{g}_{ADE}}{2},...,\overset{\mathfrak{g}_{ADE}}{2}}%
},[G_{ADE}],
\end{equation}
which we label as $\mathcal{T}_{(k)}^{\text{orb-inst}}[E_{8},G_{ADE}]$.
The heterotic description corresponds to $k$ small instantons probing a
$\mathbb{C}^2 / \Gamma_{ADE}$ singularity filled by an $E_8$ $9$-brane.

The F-theory description was worked out
in reference \cite{Aspinwall:1997ye} (see also \cite{DelZotto:2014hpa}), and the
minimal Weierstrass models are:
\begin{align}
\mathcal{T}_{(k)}^{\text{orb-inst}}[E_{8},E_{8}]  &  :y^{2}=x^{3}+s^{4}%
t^{4}x+s^{5}t^{5}(s+\alpha t^{k}) \label{E8E8orbi}\\
\mathcal{T}_{(k)}^{\text{orb-inst}}[E_{8},E_{7}]  &  :y^{2}=x^{3}+s^{4}%
t^{3}x+s^{5}t^{5}(s+\alpha t^{k})\\
\mathcal{T}_{(k)}^{\text{orb-inst}}[E_{8},E_{6}]  &  :y^{2}=x^{3}+s^{4}%
t^{3}x+s^{5}t^{4}(s+\alpha t^{k})\\
\mathcal{T}_{(k)}^{\text{orb-inst}}[E_{8},SO_{2m}]  &  :y^{2}=x^{3}+3s^{4}%
t^{2}(-1 + t^{m-4})x+2s^{5}t^{3}(s+\alpha t^{k})\\
\mathcal{T}_{(k)}^{\text{orb-inst}}[E_{8},SU_{m}]  &  :y^{2}=x^{3}%
+3s^{4}(-1+t^{m})x+2s^{5}(s+\alpha t^{k}), \label{E8SUorbi}
\end{align}
where $\alpha$ is a complex parameter which plays no role in the 6D SCFT. In the last
two lines, additional tuning is necessary in $f$ and $g$ of the Weierstrass model to realize a type $I^{\ast}_{m-4}$ and type
$I_m$ Kodaira fiber along $t = 0$.

\subsection{Fission Products}

We now demonstrate that nearly all 6D\ SCFTs can be obtained as fission
products of this simple class of progenitor theories in lines (\ref{E8E8orbi})--(\ref{E8SUorbi}).
To show this, consider a generic 6D\ SCFT on its partial tensor branch, as characterized by  (\ref{generic}).

Our primary claim is that there exists an $(E_{8},G_{\text{max}})$
orbi-instanton progenitor theory, which upon undergoing fission, yields as one
of its decay products, the theory of line (\ref{generic}).

The first step in establishing this claim is to consider possible tensor
branch flows of the rank $k$ orbi-instanton theories, with partial tensor branch
description:%
\begin{equation}
\lbrack E_{8}],\underset{k}{\underbrace{\overset{\mathfrak{g}_{\text{max}}%
}{1},\overset{\mathfrak{g}_{\text{max}}}{2},...,\overset{\mathfrak{g}%
_{\text{max}}}{2}}},[G_{\text{max}}]. \label{orbiinstagain}%
\end{equation}
This sort of blowing up procedure can either take place at the curves listed
above, or on a curve associated with 6D\ conformal matter (\ref{eq:cm}) between the listed
gauge groups on the partial tensor branch. It is
enough to consider just blowups of the $-1$ curve, as well as links on the
left and right of the quiver. If we blowup the $-1$ curve on the very left of
the diagram, we trigger a tensor branch flow to the theory of $k$ M5-branes at
the ADE\ singularity $\mathbb{C}^{2}/\Gamma_{ADE}$, which can then undergo
further Higgs branch flows. Additionally, we can instead consider a blowup of
the 6D\ conformal matter.

Since the structure of the partial tensor branch of (\ref{orbiinstagain})
has a clear repeating structure, we see that upon considering a fission process
from the orbi-instanton theories, it suffices to leave $k$ arbitrary, in which
case the number of blowups (namely the number of independent real scalars in tensor multiplets
which have non-zero vev) is either zero, one, or two.

After this, we can ask what sort of homplex deformations we can take on the
left and right sides of the resulting theory.

\begin{itemize}
	\item On the right-hand side of the
	tensor branch deformation, we have, after our tensor branch deformation, some
	choice of right flavor symmetry algebra, which we label as $\mathfrak{g}_{\mathrm{R}}$.
	The sequence of curves after the right-most copy of
	$\mathfrak{g}_\mathrm{max}$ will look like an ``incomplete'' version of one of the
	conformal matter chains in (\ref{eq:cm}). If for example $\mathfrak{g}_\mathrm{max}=\mathfrak{e}_8$,
	a possible sequence of curves would be $\ldots,\overset{\mathfrak{e}_8}{(12)},1,2,\overset{\mathfrak{sp}_{1}}{2},\overset{\mathfrak{g}_{2}}{3}%
	,1,[F_4]$; then $\mathfrak{g}_\mathrm{R}=\mathfrak{f}_4$. In this case, homplex deformations correspond to a choice of homomorphism $\mathfrak{su}(2)\rightarrow\mathfrak{f}_{4}$.
	\item On the left-hand side, the particular homplex deformation we consider will
	depend on the tensor branch deformation which preceded it. If we have retained
	the original $-1$ curve theory on the left, we need to specify the homplex
	deformation by a discrete homomorphism $\Gamma_{\text{max}}\rightarrow E_{8}$
	with $\Gamma_{\text{max}}$ the ADE\ subgroup of $SU(2)$ uniquely associated
	with the ADE\ group $G_{\text{max}}$. If, however, we have either blown up
	this $-1$ curve or any curve in the 6D conformal matter touching the $-1$
	curve, then we have an incomplete conformal matter chain just like those that can appear on the right, but written in reverse order.
	In this case we need to again specify the remnant flavor symmetry algebra
	$\mathfrak{g}_{\mathrm{L}}$ and a choice of continuous homomorphism $\mathfrak{su}%
	(2)\rightarrow\mathfrak{g}_{\mathrm{L}}$.
\end{itemize}

The possible incomplete chains along with the corresponding flavor symmetries are listed for future reference in Table \ref{tab:nilpotentgroups} below, the way they would appear at the left end. The possible chains appearing at the right end is obtained by reversing the order of the curves.

\begin{table}[ptb]
\centering
\begin{tabular}
[c]{|c|c|c|c|}\hline
$\mathfrak{g}_{\mathrm{max}}$ & $\alpha$ ($\beta^t$) & $\mathfrak{g}_{\mathrm{L}}$
($\mathfrak{g}_{\mathrm{R}}$)& incomplete chain \\\hline
\multirow{ 11}{*}{$\mf{e}_8$} & $\emptyset$ & $\mathfrak{e}_{8}$ & 12231513221\\
& 3 & $\mathfrak{f}_{4}$ & 13221\\
& 4 & $\mathfrak{g}_{2}$ & 221\\
& 5 & $\mathfrak{su}(2)$ & 21\\
& 6 & $1$ & 1 \\
& 7 & $1^{\prime}$ & $\emptyset$\\
& 23 & $\mathfrak{g}_{2}^{\prime}$ & 1513221 \\
& 33 & $1^{\prime\prime}$ & 513221 \\
& 24 & $1^{\prime\prime\prime}$ & 3221\\
& 223 & $\mathfrak{su}(2)^{\prime}$ & 31513221\\
& 2223 & $1^{\prime\prime\prime\prime}$ & 231513221 \\
& 22223 & $1^{\prime\prime\prime\prime\prime}$ & 2231513221\\\hline
\multirow{ 6}{*}{$\mf{e}_7$} & $\emptyset$ & $\mathfrak{e}_{7}$ & 12321\\
& 3 & $\mathfrak{so}(7)$ & 21 \\
& 4 & $\mathfrak{su}(2)$ & 1\\
& 5 & $1$ & $\emptyset$\\
& 23 & $\mathfrak{su}(2)^{\prime}$ & 321\\
& 223 & $1^{\prime}$ & 2321 \\\hline
\multirow{ 4}{*}{$\mathfrak{e}_6$} & $\emptyset$ & $\mathfrak{e}_{6}$ & 131\\
& 3 & $\mathfrak{su}(3)$ & 1\\
& 4 & $1$ & $\emptyset$ \\
& 23 & $1^{\prime}$ & 31 \\\hline
\multirow{ 2}{*}{$\mf{so}(2k)$} & $\emptyset$ & $\mathfrak{so}(2k)$ & 1 \\
& $3$ & $\mathfrak{sp}(k-4)$ & $\emptyset$ \\\hline
{$\mathfrak{su}(k)$} & $\emptyset$ & $\mathfrak{su}(k)$ & $\emptyset$\\\hline
\end{tabular}
\caption{Choices of $\mathfrak{g}_{\mathrm{L}}$ ($\mathfrak{g}_{\mathrm{R}}$) as a function of
$\mathfrak{g}_{\mathrm{max}}$ and endpoint $\alpha$ ($\beta^t$), where the superscript $t$ indicates
that we reverse the order of the curves by transposition. We also tabulate
the corresponding incomplete chains (or their transpose)
appearing after the leftmost (rightmost) $\mathfrak{g}_\mathrm{max}$.
}%
\label{tab:nilpotentgroups}%
\end{table}

A nontrivial fact, checked in \cite{Mekareeya:2017sqh}, is that after the first stage of tensor deformations,
the list of theories we obtain covers all the possible A-type endpoints,
which have been classified in \cite{Heckman:2013pva}.

With these encouraging preliminaries in mind, we have checked by inspection of the resulting quiver-like theories
obtained in reference \cite{Heckman:2015bfa} that the vast majority of 6D\ SCFTs can
therefore alternatively be labeled through the following steps:

\begin{enumerate}
\item Select an ADE-type gauge algebra $\mathfrak{g}_{\mathrm{max}}$.

\item Select $\ell_{\text{plat}}\geq1$, the number of times $\mathfrak{g}%
_{\mathrm{max}}$ appears in the quiver.

\item For the given $\mathfrak{g}_{\mathrm{max}}$, select some $\mathfrak{g}%
_{\mathrm{R}}$ from (\ref{eq:cm}) and an associated nilpotent
orbit $\mathcal{O}_{\mathrm{R}}$ of $\mathfrak{g}_{\mathrm{R}}$.

\item For the given $\mathfrak{g}_{\mathrm{max}}$, select \emph{either} some
$\mathfrak{g}_{\mathrm{L}}$ from (\ref{eq:cm}) and an
associated nilpotent orbit $\mathcal{O}_{\mathrm{L}}$ of $\mathfrak{g}_{\mathrm{L}}$ \emph{or} a
homomorphism in $\Gamma_{\mathfrak{g}_{\mathrm{max}}}\rightarrow E_{8}$ (if
the flow from the progenitor theory leaves intact the leftmost $-1$ curve of
line (\ref{orbiinstagain})).
\end{enumerate}

We remark that in Step 4, the resulting homplex deformation naturally splits
into two cases, based on whether or not the blown down $-1$ curve on the partial tensor
branch remains as part of the 6D\ SCFT.\ In the
case where this $-1$ curve is no longer part of the 6D\ SCFT, the new endpoint
for the theory obtained by successively blowing down all $-1$ curves is
necessarily non-trivial, and the singular base of the resulting SCFT is an
orbifold $\mathbb{C}^{2}/\Gamma_{U(2)}$ with $\Gamma_{U(2)}$ a non-trivial
finite subgroup of $U(2)$. This leads to some additional refinements in the
resulting fission products which can emerge, which we now describe.

\subsubsection{Refinements with a Long A-type Endpoint}

Consider then, the theories with a non-trivial long A-type endpoint.
Recall that these are labeled by a sequence of $\ell_{\text{end}}$ integers.
For sufficiently large $\ell_{\text{end}}$, these take the form
\begin{equation}\label{eq:Anendpoint}
\alpha22...22\beta
\end{equation}
Where $\alpha$, $\beta$ are restricted to be one of the following \cite{Heckman:2013pva}:
\begin{align}\label{eq:alpha-beta}
\alpha &  \in\{\emptyset,3,4,5,6,7,23,33,24,223,2223,22223\}\nonumber\\
\beta &  \in\{\emptyset,3,4,5,6,7,32,33,42,322,3222,32222\}
\end{align}
Here, $\emptyset$ indicates that $\alpha$ or $\beta$ may be trivial, as in the
case of $(2,0)$ SCFTs, or the worldvolume theory of a stack of M5-branes
probing a $\mathbb{C}^{2}/\Gamma_{ADE}$ orbifold singularity with
$\Gamma_{ADE}\subset SU(2)$ a finite subgroup.

For sufficiently many curves in the endpoint, namely for $\ell_{\text{end}}$, such
theories exhibit the \textquotedblleft generic\textquotedblright\ behavior of 6D\ SCFTs
\cite{Heckman:2013pva, Heckman:2015bfa}. Theories with a D- or E-type endpoint, as well as models
with shorter endpoints can exhibit ``outlier'' behavior. We analyze short bases which
fit into the general pattern of fission products in subsection \ref{ssec:short copy(1)}
and explain how all remaining outliers are generated via fusion in section \ref{sec:DARTHFUSIUS}.

Much as in the more general case where we start from our progenitor theories, we can label most 6D\ SCFTs via three steps:
\begin{enumerate}
\item Select an ADE-type gauge algebra $\mathfrak{g}_{\mathrm{max}}$.

\item Select an A-type \textquotedblleft endpoint\textquotedblright\ configuration.

\item Select a pair of nilpotent orbits $\mathcal{O}_{\mathrm{L}}$, $\mathcal{O}_{\mathrm{R}}$
of $\mathfrak{g}_{\mathrm{L}}$, $\mathfrak{g}_{\mathrm{R}}$, respectively.
\end{enumerate}

As previously mentioned, after the first step of tensor branch deformation (before homplex deformations) one already covers all the possible A-type endpoints. In fact, it was found in \cite{Mekareeya:2017sqh} that this is true even without considering the theories with $\Gamma_{\text{max}}\rightarrow E_{8}$ homomorphisms; there is a one-to-one correspondence between the set of theories with incomplete conformal matter chains on both sides (before homplex deformations) and the set of the possible endpoints found in \cite{Heckman:2013pva}. For long endpoints, this covers in particular all the choices allowed in line (\ref{eq:alpha-beta}); the one-to-one correspondence is expressed by the second and fourth columns of Table \ref{tab:nilpotentgroups}. The correspondence is still valid for short endpoints: even the outliers found in \cite{Heckman:2013pva} are reproduced by that table with a simple formal rule, which we will see in section \ref{ssec:short copy(1)}. However, for outlier endpoints homplex deformations with nilpotent orbits on both sides fail to produce all possible theories, whereas for long enough endpoints they do produce all theories.

Additionally, we note that complex structure deformations cannot change the endpoint. So
in particular, homplex deformations do not affect the endpoint. This implies that fission
reproduces all the possible endpoints.

Given now an endpoint in (\ref{eq:Anendpoint}), we associate a gauge algebra to each of the
numbers in the sequence. The allowed ways of doing this were classified in
\cite{Heckman:2015bfa}. As we saw in (\ref{generic}), one of the main punchlines of that analysis is that
these gauge algebras obey a ``convexity condition," increasing as one moves
from the outside in and reaching a maximum somewhere in the interior of the
sequence. In the present classification, $\mathfrak{g}_{\mathrm{max}}$ is
defined to be the largest gauge algebra. One then arrives at a 6D SCFT quiver by decorating the above sequence with additional ``links." For large $\ell_{\text{end}}$, the quiver is uniquely fixed in the
interior, and ambiguities arise only at the far left and far right. Thus, one
of these 6D SCFTs with sufficiently large $\ell_{\text{end}}$ is labeled by a choice of
endpoint, a maximal gauge algebra $\mathfrak{g}%
_{\mathrm{max}}$, and a pair of decorations, one on the far left and on the
far right.

These decorations are classified by nilpotent orbits of
gauge algebras \cite{Heckman:2016ssk}. In the case that $\alpha$ ($\beta$) is
trivial, decorations on the far left (right) of the quiver are labeled simply
by nilpotent orbits of $\mathfrak{g}_{\mathrm{max}}$.
This case was analyzed at length in \cite{Heckman:2016ssk}, where the one-to-one correspondence was shown explicitly for all nilpotent orbits in any $\mathfrak{g}_\mathrm{max}$.

For more general endpoints, these decorations are labeled by nilpotent orbits of some
subalgebra $\mathfrak{g}_{\mathrm{L}}, \mathfrak{g}_{\mathrm{R}} \subset\mathfrak{g}_{\mathrm{max}}$.
Table \ref{tab:nilpotentgroups} shows the correspondence between endpoints and subalgebras.

We have also checked that the Higgs moduli spaces of homplex deformations obey the simple rule obtained in \cite{Mekareeya:2016yal} for chains of conformal matter theories: namely, that the difference in Higgs moduli space dimensions $d_\mathrm{H}$ between the deformed and original theory is given by
\begin{equation}
	\Delta d_\mathrm{H} = \mathrm{dim} {\cal O}_\mathrm{L} + \mathrm{dim} {\cal O}_\mathrm{R}\,.
\end{equation}
Here, the left-hand side can be computed via its relation to a coefficient in the anomaly polynomial of the 6D SCFT.

\subsection{Examples}\label{ssec:examples}

It is helpful to illustrate the above considerations with some explicit
examples. This also shows how non-trivial it is for nearly all 6D\ SCFTs to
descend from such a small class of progenitor theories. Strictly speaking, we
have already established that the primary progenitors are the orbi-instanton
theories, in which case we need to further distinguish between tensor branch
flows which retain the leftmost $-1$ curve of (\ref{orbiinstagain}) and
those which do not, as this dictates the kind of homplex deformation we are
dealing with. Though technically redundant, it is helpful to also consider
separately the fission products from 6D conformal matter theories i.e.
theories of M5-branes probing an ADE singularity. We now turn to examples of
each type.

\subsubsection{Fission from the Orbi-Instanton Theories}

First, let us consider the case of
$\mathfrak{g}_{\mathrm{max}}=\mathfrak{su}(3)$, $\ell_{\text{plat}}=5$,
$\mathfrak{g}_{\mathrm{R}}=\mathfrak{su}(3)$, $\mathcal{O}_{\mathrm{R}}=[2,1]$. On the left,
consider a homomorphism $\mathbb{Z}_{3}\rightarrow E_{8}$ obtained by deleting
the third node of the affine $E_{8}$ Dynkin diagram. The algorithm of Kac detailed in
reference \cite{MR739850} tells us the unbroken symmetry group, and this was applied in the
context of 6D SCFTs in reference \cite{Heckman:2015bfa}.
This homomorphism leaves unbroken a subalgebra $\mathfrak{e}_{6}%
\times\mathfrak{su}(3)\subset\mathfrak{e}_{8}$, and the resulting 6D SCFT
quiver is:
\[
\lbrack E_{6}]\,\,1\,\,\underset{[SU(3)]}{\overset{\mathfrak{su}(3)}{2}%
}\,\,\overset{\mathfrak{su}(3)}{2}\,\,\overset{\mathfrak{su}(3)}{2}%
\,\,\overset{\mathfrak{su}(3)}{2}\,\,\overset{\mathfrak{su}(3)}{2}%
\,\,\underset{[N_{f}=1]}{\overset{\mathfrak{su}(2)}{2}}%
\]
Note that there are five $\mathfrak{su}(3)$ gauge algebras, corresponding to
$\ell_{\text{plat}}=5$.

As a second example, let us consider select the theory with $\mathfrak{g}%
_{\mathrm{max}}=\mathfrak{e}_{6}$, $\ell_{\text{plat}}=3$, $\mathfrak{g}%
_{\mathrm{R}}=1^{\prime}$ (forcing $\mathcal{O}_{\mathrm{R}}$ to be trivial), $\mathfrak{g}%
_{\mathrm{L}}=\mathfrak{e}_{6}$, and $\mathcal{O}_{\mathrm{L}}=A_{2}+2A_{1}$, (labeling a nilpotent orbit
by its associated Bala--Carter label). This corresponds to the theory:
\begin{equation}
\lbrack U(2)]\,\,\overset{\mathfrak{e}_{6}}{4}\,\,1\,\,\overset{\mathfrak{su}%
(3)}{3}\,\,1\,\,\,\,\overset{\mathfrak{e}_{6}}{6}%
\,\,1\,\,\overset{\mathfrak{su}(3)}{3}\,\,1\,\,\,\,\overset{\mathfrak{e}%
_{6}}{6}\,\,1\,\,\overset{\mathfrak{su}(3)}{3}%
\end{equation}
There are three $\mathfrak{e}_{6}$ gauge algebras, corresponding to
$\ell_{\text{plat}}=3$. On the left, the flavor symmetry is indeed $U(2)$,
which is the subgroup of $E_{6}$ left unbroken by the nilpotent orbit
$A_{2}+2A_{1}$.


\subsubsection{Fission from 6D\ Conformal Matter}

Consider next some examples involving flows from the theories with just $-2$
curves. As a first example, we consider theories with $\mathfrak{g}%
_{\mathrm{max}}=\mathfrak{su}(m)$. Such theories necessarily have
$\alpha=\beta=\emptyset$, so we begin with a quiver of the form:
\begin{equation}
\lbrack\mathfrak{su}(m)]\,\,\underset{k}{\underbrace{\overset{\mathfrak{su}%
(m)}{2}\,\,\overset{\mathfrak{su}(m)}{2}\,\,\cdots\,\,\overset{\mathfrak{su}%
(m)}{2}}}\,\,[\mathfrak{su}(m)]
\end{equation}
Here, every curve in the endpoint is associated with a $\mathfrak{su}(m)$
gauge algebra. The intermediate links between neighboring gauge algebras are
simply bifundamentals $(\mathbf{m},\overline{\mathbf{m}})$, and there are
flavor symmetries $\mathfrak{su}(m)_{\mathrm{L}}$ and $\mathfrak{su}(m)_{\mathrm{R}}$ on the far
left and right, respectively. This is the quiver for the worldvolume theory of
$k+1$ M5-branes probing a $\mathbb{C}^{2}/\mathbb{Z}_{m}$ orbifold singularity.

We may deform this quiver at the far left and far right by nilpotent orbits of
$\mathfrak{su}(m)$. Such nilpotent orbits are labeled simply by partitions of
$m$. The dictionary between partitions and deformations of the quiver is as
follows: given a pair of partitions $\mu_{\mathrm{L}}$, $\mu_{\mathrm{R}}$ and labeling the
gauge algebras from left to right as $\mathfrak{su}(m_{1}),...,\mathfrak{su}%
(m_{k})$:
\begin{equation}
\underset{\ell_{L}}{\underbrace{\overset{\mathfrak{su}\left(  m_{1}\right)
}{2},...,\overset{\mathfrak{su}\left(  m_{\ell_{L}}\right)  }{2}}%
},\underset{\ell_{\text{plat}}}{\underbrace{\overset{\mathfrak{su}\left(
m_{\text{max}}\right)  }{2},...,\overset{\mathfrak{su}\left(  m_{\text{max}%
}\right)  }{2}}},\underset{\ell_{R}}{\underbrace{\overset{\mathfrak{su}\left(
m_{k-\ell_{R}+1}\right)  }{2},...,\overset{\mathfrak{su}\left(  m_{k}\right)
}{2}}}
\end{equation}
where in the above, we suppress the flavor symmetry factors. Then, the ramps
on the left and right obey:
\begin{equation}
\begin{split}
	m_{1}  &  =(\mu_{\mathrm{L}}^{T})_{1},~~m_{2}=(\mu_{\mathrm{L}}^{T})_{1}+(\mu_{\mathrm{L}}^{T}%
	)_{2},~~...~m_{\ell_{L}}=\sum_{i=1}^{\ell_{L}}(\mu_{\mathrm{L}}^{T})_{i}\nonumber\\
	m_{k}  &  =(\mu_{\mathrm{R}}^{T})_{1},~~m_{k-1}=(\mu_{\mathrm{R}}^{T})_{1}+(\mu_{\mathrm{R}}^{T}%
	)_{2},~~...~m_{k-\ell_{R} + 1}=\sum_{i=1}^{\ell_{R}}(\mu_{\mathrm{R}}^{T})_{i}. \label{eq:keq}%
\end{split}	
\end{equation}
In this case, the statement that the quiver is \textquotedblleft sufficiently
long," means it should be long enough so that the deformations on the left and
right are separated by a plateau in which $m_{i}=m$.

As a slightly more involved example, we consider the
endpoint (\ref{eq:Anendpoint}) with $\alpha=3$, $\beta=\emptyset$:
\begin{equation}
3 \, 2 \, 2\cdots\, 2\,.
\end{equation}
We see from Table \ref{tab:nilpotentgroups} that both $\alpha=3$ and $\beta=\emptyset$ can occur for any $\mathfrak{g}_\mathrm{max}\neq \mathfrak{su}$; let us pick $\mathfrak{g}_{\mathrm{max}} = \mathfrak{e}_{6}$.
In this case, since $\mathfrak{g}_{\mathrm{max}} = \mathfrak{e}_{6}$, we must
resolve the F-theory base to move to the full tensor branch of the theory,
introducing intermediate conformal matter links (recall (\ref{eq:cm}))
between the $\mathfrak{e}_{6}$ gauge algebras. A full resolution yields
\begin{equation}
[\mathfrak{su}(3)] \,\,1 \,\,\overset{\mathfrak{e}_{6}}{6 }\,\,1 \,\,
\overset{\mathfrak{su}(3)}{3 }\,\, 1 \,\,\cdots\,\,\overset{\mathfrak{e}%
_{6}}{6}\,\, 1 \,\, \overset{\mathfrak{su}(3)}{3 }\,\, 1 \,\, [\mathfrak{e}%
_{6}]
\end{equation}
in agreement with the last column in Table \ref{tab:nilpotentgroups}.

We want to consider deformations of this quiver which preserve the endpoint as
well as the $\mathfrak{g}_{\mathrm{max}}=\mathfrak{e}_{6}$ plateau, but which
modify the quiver on the far left and far right. On the
far left, such deformations are also labeled by nilpotent orbits of the flavor
symmetry, $\mathfrak{su}(3)$. Explicitly, we have two possible deformations on the left,
corresponding to partitions $[2,1]$ and $[3]$, respectively:
\begin{subequations}\label{eq:32...2leftdef}
\begin{align}
\underset{[N_{f}=1]}{\overset{\mathfrak{e}_{6}}{5}} \,\,1 \,\,
\overset{\mathfrak{su}(3)}{3 }\,\, 1 \,\,\cdots\,\,\overset{\mathfrak{e}%
_{6}}{6}\,\, 1 \,\, \overset{\mathfrak{su}(3)}{3 }\,\, 1 \,\, [\mathfrak{e}%
_{6}]\\
\overset{\mathfrak{f}_{4}}{5 }\,\,1 \,\, \overset{\mathfrak{su}(3)}{3 }\,\, 1
\,\,\cdots\,\,\overset{\mathfrak{e}_{6}}{6}\,\, 1 \,\, \overset{\mathfrak{su}%
(3)}{3 }\,\, 1 \,\, [\mathfrak{e}_{6}]
\end{align}	
\end{subequations}
On the far right, since
$\beta= \emptyset$, Table \ref{tab:nilpotentgroups} tells us that such
deformations are labeled by nilpotent orbits of the flavor symmetry
$\mathfrak{e}_{6}$, as shown in the appendix of \cite{Heckman:2016ssk}.

Activating these further homplex deformations gives rise for both (\ref{eq:32...2leftdef}) to a hierarchy of possibilities,
each with a partial ordering in one-to-one correspondence with the partial ordering of $\mathfrak{e}_6$ nilpotent orbits.

\subsection{Short Bases \label{ssec:short copy(1)}}

The above considerations cover the generic behavior of
6D\ SCFTs, which in particular covers all long bases with an A-type endpoint.
Even for short bases (i.e.~those with a tensor branch of low rank), some of
the cases can be accommodated through extension of patterns observed with
higher rank tensor branches. We define a \textquotedblleft short
base\textquotedblright\ as one in which we have a non-trivial endpoint
configuration with $1\leq\ell_{\text{end}}\leq9$.

In section 5 of \cite{Heckman:2013pva}, a handful of apparent
\textquotedblleft outlying" endpoints were identified in which the number of
curves is less than or equal to nine. For instance, the endpoint $(12)$
clearly does not fit the pattern of (\ref{eq:Anendpoint}). However, as
observed in Appendix A of \cite{Morrison:2016nrt}, even these apparent
outliers can be viewed as limits of endpoints in (\ref{eq:Anendpoint}). For
instance, a single $-12$ curve is the formal limit of the endpoint $\underset{\ell_{\text{end}%
}}{\underbrace{722...27}}$ with $\ell_{\text{end}}\rightarrow1$. To
see this, we add $\mathfrak{e}_{8}$ gauge algebras and resolve the geometry to
move to the full tensor branch of the theory. The endpoint $(12)$ gives
simply
\begin{equation}
\overset{\mathfrak{e}_{8}}{(12)}. \label{eq:single12}%
\end{equation}
whereas the endpoint $722...27$ blows up to
\begin{equation}\label{eq:722...27}
\overset{\mathfrak{e}_{8}}{(12)}\,\,1\,\,2\,\,\overset{\mathfrak{su}%
(2)}{2}\,\,\overset{\mathfrak{g}_{2}}{3}\,\,1\,\,\overset{\mathfrak{f}_{4}%
}{5}\,\,1\,\,\overset{\mathfrak{g}_{2}}{3}\,\,\overset{\mathfrak{su}%
(2)}{2}\,\,2\,\,1\,\,\overset{\mathfrak{e}_{8}}{(12)}\,\,\cdots\,\,\overset{\mathfrak{e}%
_{8}}{(12)}\,\,1\,\,2\,\,\overset{\mathfrak{su}(2)}{2}\,\,\overset{\mathfrak{g}_{2}%
}{3}\,\,1\,\,\overset{\mathfrak{f}_{4}}{5}\,\,1\,\,\overset{\mathfrak{g}%
_{2}}{3}\,\,\overset{\mathfrak{su}(2)}{2}\,\,2\,\,1\,\,\overset{\mathfrak{e}%
_{8}}{(12)}%
\end{equation}
We see that this reduces to (\ref{eq:single12}) in the limit in which the
number of $\mathfrak{e}_{8}$ gauge algebras goes to 1. The rest of the
apparent \textquotedblleft outliers" behave similarly: in this sense, there
are no outlier endpoints.

The general rule to obtain short endpoints from long ones is as follows \cite{Mekareeya:2017sqh}. One should think of outlier endpoints as obtained by ``analytically continuing'' the number of $2$'s in (\ref{eq:Anendpoint}) to $-1$, and by applying the operation $\ldots x 2^{-1} y \ldots \mapsto \ldots (x+y-2)\ldots$. This rule reproduces Table 2 of reference \cite{Morrison:2016nrt}.
For example, the outlier endpoint $22228$ is obtained by $\alpha 2^{-1} \beta$ for $\alpha=22223$ and $\beta=7$.
As another example, for $\alpha=7$ and $\beta=7$ we obtain $72^{-1}7\mapsto (7+7-2)=(12)$,
in agreement with the example (\ref{eq:single12})--(\ref{eq:722...27}) above.

Indeed, most theories at small $\ell_{\text{end}}$ can in fact be given a group-theoretic
description as above, with nilpotent orbits $\mathcal{O}_{\mathrm{L}}$, $\mathcal{O}%
_{\mathrm{R}}$ overlapping in a nontrivial way. The story is simplest in the case of
$\mathfrak{g}_{\mathrm{max}} = \mathfrak{su}(m)$. For concreteness, let us
consider the case of $\ell_{\text{end}}=3$, so the quiver takes the form
\begin{equation}
\underset{[\mathfrak{su}(n_1)]}{\overset{\mathfrak{su}(m_1)}{2}}
\,\,\underset{[\mathfrak{su}(n_2)]}{\overset{\mathfrak{su}(m_2)}{2}}
\,\,\underset{[\mathfrak{su}(n_3)]}{\overset{\mathfrak{su}(m_3)}{2}}%
\end{equation}
where max$(m_{i}) = m$. Here, the $[\mathfrak{su}(n_i)]$ denote
flavor symmetries chosen so as to cancel all gauge anomalies.
As before, such theories are labeled by a pair of
nilpotent orbits of $\mathfrak{su}(m)$, and hence partitions $\mu_{\mathrm{L}}$,
$\mu_{\mathrm{R}}$ of $m$. Now, however, there is a constraint on these partitions: the
total number of rows in the pair of partitions must be less than or equal to
$\ell_{\text{end}}+1=4$. To see why, let us take $m=5$ and try to set $\mu_{\mathrm{L}}=[5]$, which has
5 rows. From (\ref{eq:keq}), we then have $m_{1}=1$, $m_{2}=2$, $m_{3}=3$,
which means that max$(m_{i}) \neq m$! One might say that this particular
nilpotent orbit has ``run out of room:'' it requires a longer quiver because it
induces a deformation of the quiver far into the interior. On the other hand,
one may take e.g.~$\mu_{\mathrm{L}}=[2^{2},1],\ \mu_{\mathrm{R}}=[2,1^{3}]$ since the total
number of rows between these two partitions is four, leading to the quiver
\begin{equation}
\underset{[N_{f}=1]}{\overset{\mathfrak{su}(3)}{2}}
\,\,\underset{[\mathfrak{su}(3)]}{\overset{\mathfrak{su}(5)}{2}}
\,\,\underset{[\mathfrak{su}(3)]}{\overset{\mathfrak{su}(4)}{2}}%
\end{equation}

For the case of exceptional $\mathfrak{g}_{\mathrm{max}}$, pairs of nilpotent
orbits can be used to produce some rather exotic theories at small $\ell_{\text{end}}$,
including those for which all of the $\mathfrak{g}_{\mathrm{max}}$ algebras
are Higgsed to a subalgebra. For instance, the quiver
\begin{equation}
1\,\,\overset{\mathfrak{f}_{4}}{5}\,\,1\,\,\overset{\mathfrak{g}_{2}%
}{3}\,\,\overset{\mathfrak{sp}(1)}{1} \label{eq:G2SO13ex}%
\end{equation}
does not look like it fits in with the group-theoretic classification that
worked at large $\ell_{\text{end}}$, but in fact it may be realized as the $\ell_{\text{end}}=1$ limit of the
theory with $\mathfrak{g}_{\mathrm{max}}=\mathfrak{e}_{7}$, $\mathcal{O}%
_{\mathrm{L}}=A_{5}^{\prime\prime}$, $\mathcal{O}_{\mathrm{R}}=A_{1}$ (again, labeling nilpotent
orbits by their Bala--Carter labels):
\begin{equation}
1\,\,\overset{\mathfrak{f}_{4}}{5}\,\,1\,\,\overset{\mathfrak{g}_{2}%
}{3}\,\,\overset{\mathfrak{su}(2)}{2}\,\,1\,\,\overset{\mathfrak{e}_{7}%
}{8}\,\,\cdots\,\,\overset{\mathfrak{e}_{7}}{8}\,\,1\,\,\overset{\mathfrak{su}%
(2)}{2}\,\,\overset{\mathfrak{so}(7)}{3}\,\,\overset{\mathfrak{sp}(1)}{1}
\label{eq:G2SO12ex}%
\end{equation}
This can be verified by computing the anomaly polynomial of the class of
theories with $\mathfrak{g}_{\mathrm{max}}=\mathfrak{e}_{7}$, $\mathcal{O}%
_{\mathrm{L}}=A_{5}^{\prime\prime}$, $\mathcal{O}_{\mathrm{R}}=A_{1}$ as a function of $\ell_{\text{end}}$ using
the prescription of \cite{Ohmori:2014kda} and analytically continuing to
$\ell_{\text{end}}=1$.

By taking limits of theories labeled by a pair of homomorphisms, one can
produce a large class of 6D SCFTs with small $\ell_{\text{end}}$. Along these
lines, we now show how all of the ``non-Higgsable clusters" (NHCs) of \cite{Morrison:2012np}
arise in this way:
\begin{align}
\overset{\mathfrak{su}(3)}{3}  &  = \lim_{\ell_{\text{plat}} \rightarrow0}
\overset{\mathfrak{su}(3)}{3 }\,\, 1 \,\, \overset{\mathfrak{su}(3)}{3 }\,\, 1
\,\, \overset{\mathfrak{e}_{6}}{6 }\,\, \cdots\,\,1 \,\,
\overset{\mathfrak{su}(3)}{3 }\,\, 1 \,\, \overset{\mathfrak{e}_{6}%
}{6}\nonumber\\
&  \Rightarrow\mathfrak{g}_{\mathrm{max}} = \mathfrak{e}_{6},\ell_{\text{plat}} =0,
\mathcal{O}_{\mathrm{L}} \in\text{Hom}(\Gamma_{E_{6}} , E_{8}), \mathfrak{g}_{\mathrm{R}} =
1\label{eq:firstNHC}\\
\overset{\mathfrak{so}(8)}{4 }  &  = \lim_{\ell_{\text{end}} \rightarrow1}
\overset{\mathfrak{so}(8)}{4 }\,\, 1 \,\, \overset{\mathfrak{so}(8)}{4 }\,\, 1
\,\, \cdots\,\, \overset{\mathfrak{so}(8)}{4 }\,\, 1 \,\,
\overset{\mathfrak{so}(8)}{4}\nonumber\\
&  \Rightarrow\mathfrak{g}_{\mathrm{max}} = \mathfrak{so}(8), \ell_{\text{end}} =1,
\mathfrak{g}_{\mathrm{L}} = 1, \mathfrak{g}_{\mathrm{R}} = 1\\
\overset{\mathfrak{f}_{4}}{5 }  &  = \lim_{\ell_{\text{end}} \rightarrow1}
\overset{\mathfrak{f}_{4}}{5 }\,\, 1 \,\, \overset{\mathfrak{su}(3)}{3 }\,\, 1
\,\, \overset{\mathfrak{e}_{6}}{6 }\,\, \cdots\,\,1 \,\,
\overset{\mathfrak{su}(3)}{3 }\,\, 1 \,\, \overset{\mathfrak{e}_{6}%
}{6}\nonumber\\
&  \Rightarrow\mathfrak{g}_{\mathrm{max}} = \mathfrak{e}_{6}, \ell_{\text{end}} =1,
\mathfrak{g}_{\mathrm{L}} = \mathfrak{su}(3), \mathcal{O}_{\mathrm{L}} = [3], \mathfrak{g}_{\mathrm{R}} =
1\\
\overset{\mathfrak{e}_{6}}{6 }  &  = \lim_{\ell_{\text{end}} \rightarrow1}
\overset{\mathfrak{e}_{6}}{6 }\,\, 1 \,\, \overset{\mathfrak{su}(3)}{3 }\,\, 1
\,\, \overset{\mathfrak{e}_{6}}{6 }\,\, \cdots\,\,1 \,\,
\overset{\mathfrak{su}(3)}{3 }\,\, 1 \,\, \overset{\mathfrak{e}_{6}%
}{6}\nonumber\\
&  \Rightarrow\mathfrak{g}_{\mathrm{max}} = \mathfrak{e}_{6}, \ell_{\text{end}} =1,
\mathfrak{g}_{\mathrm{L}} = 1, \mathfrak{g}_{\mathrm{R}} = 1\\
\underset{[N_{f}=1/2]}{\overset{\mathfrak{e}_{7}}{7}}  &  = \lim_{\ell_{\text{end}}
\rightarrow1} \underset{[N_{f}=1/2]}{\overset{\mathfrak{e}_{7}}{7}} \,\, 1
\,\, \overset{\mathfrak{su}(2)}{2 }\,\, \overset{\mathfrak{so}(7)}{3 }\,\,
\overset{\mathfrak{su}(2)}{2 }\,\, 1 \,\, \overset{\mathfrak{e}_{7}}{8 }\,\,
\cdots\,\, \overset{\mathfrak{so}(7)}{3 }\,\, \overset{\mathfrak{su}(2)}{2
}\,\, 1 \,\, \overset{\mathfrak{e}_{7}}{8}\nonumber\\
&  \Rightarrow\mathfrak{g}_{\mathrm{max}} = \mathfrak{e}_{7}, \ell_{\text{end}} =1,
\mathfrak{g}_{\mathrm{L}} = \mathfrak{su}(2), \mathcal{O}_{\mathrm{L}} = [2], \mathfrak{g}_{\mathrm{R}} =
1\\
{\overset{\mathfrak{e}_{7}}{8}}  &  = \lim_{\ell_{\text{end}} \rightarrow1}
{\overset{\mathfrak{e}_{7}}{8}} \,\, 1 \,\, \overset{\mathfrak{su}(2)}{2 }\,\,
\overset{\mathfrak{so}(7)}{3 }\,\, \overset{\mathfrak{su}(2)}{2 }\,\, 1 \,\,
\overset{\mathfrak{e}_{7}}{8 }\,\, \cdots\,\, \overset{\mathfrak{so}(7)}{3
}\,\, \overset{\mathfrak{su}(2)}{2 }\,\, 1 \,\, \overset{\mathfrak{e}_{7}%
}{8}\nonumber\\
&  \Rightarrow\mathfrak{g}_{\mathrm{max}} = \mathfrak{e}_{8}, \ell_{\text{end}} =1,
\mathfrak{g}_{\mathrm{L}} = 1, \mathfrak{g}_{\mathrm{R}} = 1\\
\overset{\mathfrak{e}_{8}}{(12)}  &  = \lim_{\ell_{\text{end}} \rightarrow1}
\overset{\mathfrak{e}_{8}}{(12)} \,\, 1 \,\, 2 \,\, \overset{\mathfrak{su}%
(2)}{2 }\,\, \overset{\mathfrak{g}_{2}}{3 }\,\, \cdots\,\,
\overset{\mathfrak{g}_{2}}{3 }\,\, \overset{\mathfrak{su}(2)}{2 }\,\, 2 \,\, 1
\,\, \overset{\mathfrak{e}_{8}}{(12)}\nonumber\\
&  \Rightarrow\mathfrak{g}_{\mathrm{max}} = \mathfrak{e}_{8}, \ell_{\text{end}} =1,
\mathfrak{g}_{\mathrm{L}} =1^{\prime}, \mathfrak{g}_{\mathrm{R}} = 1^{\prime}\\
\overset{\mathfrak{g}_{2}}{3 }\,\, \overset{\mathfrak{su}(2)}{2 }  &  =
\lim_{\ell_{\text{end}} \rightarrow2} \,\, \overset{\mathfrak{g}_{2}}{3 }\,\,
\overset{\mathfrak{su}(2)}{2 }\,\, 1 \,\, \overset{\mathfrak{e}_{6}}{6 }\,\,1
\,\, \overset{\mathfrak{su}(3)}{3 }\,\, 1 \,\, \cdots\,\, 1 \,\,
\overset{\mathfrak{su}(3)}{3 }\,\, 1 \,\, \overset{\mathfrak{e}_{6}}{6 }\,\, 1
\,\, \overset{\mathfrak{su}(3)}{3}\nonumber\\
&  \Rightarrow\mathfrak{g}_{\mathrm{max}} = \mathfrak{e}_{6}, \ell_{\text{end}} =2,
\mathfrak{g}_{\mathrm{L}} = \mathfrak{e}_{6}, \mathcal{O}_{\mathrm{L}} = A_{4}+A_{1},
\mathfrak{g}_{\mathrm{R}} = 1^{\prime}\\
\overset{\mathfrak{g}_{2}}{3 }\,\, \overset{\mathfrak{su}(2)}{2 }\,\, 2  &  =
\lim_{\ell_{\text{end}} \rightarrow3} \,\, \overset{\mathfrak{g}_{2}}{3 }\,\,
\overset{\mathfrak{su}(2)}{2 }\,\, 2 \,\, 1 \,\, {\overset{\mathfrak{e}%
_{7}}{8}} \,\, 1 \,\, \overset{\mathfrak{su}(2)}{2 }\,\,
\overset{\mathfrak{so}(7)}{3 }\,\, \cdots\,\, \overset{\mathfrak{su}(2)}{2
}\,\, 1 \,\, \overset{\mathfrak{e}_{7}}{8 }\,\, 1 \,\, \overset{\mathfrak{su}%
(2)}{2 }\,\, \overset{\mathfrak{so}(7)}{3 }\,\, \overset{\mathfrak{su}%
(2)}{2}\nonumber\\
&  \Rightarrow\mathfrak{g}_{\mathrm{max}} = \mathfrak{e}_{7}, \ell_{\text{end}} =3,
\mathfrak{g}_{\mathrm{L}} = \mathfrak{e}_{7}, \mathcal{O}_{\mathrm{L}} = D_{6}(a_{1}),
\mathfrak{g}_{\mathrm{R}} = 1^{\prime}\\
\overset{\mathfrak{su}(2)}{2 }\,\, \overset{\mathfrak{so}(7)}{3 }\,\,
\overset{\mathfrak{su}(2)}{2 }  &  = \lim_{\ell_{\text{end}} \rightarrow3}
\,\,\overset{\mathfrak{su}(2)}{2 }\ \,\,\overset{\mathfrak{so}(7)}{3 }\,\,
\overset{\mathfrak{su}(2)}{2 }\,\, 1 \,\, \overset{\mathfrak{e}_{6}}{6 }\,\,1
\,\, \overset{\mathfrak{su}(3)}{3 }\,\, 1 \,\, \cdots\,\, 1 \,\,
\overset{\mathfrak{su}(3)}{3 }\,\, 1 \,\, \overset{\mathfrak{e}_{6}}{6 }\,\, 1
\,\, \overset{\mathfrak{su}(3)}{3}\nonumber\\
&  \Rightarrow\mathfrak{g}_{\mathrm{max}} = \mathfrak{e}_{6}, \ell_{\text{end}} =3,
\mathfrak{g}_{\mathrm{L}} = \mathfrak{e}_{6}, \mathcal{O}_{\mathrm{L}} = D_{5}, \mathfrak{g}_{\mathrm{R}}
= 1^{\prime} \label{eq:lastNHC}%
\end{align}
Some of these limits are quite obvious, especially those that produce the
$-4$, $-6$, $-8$, and $-12$ NHCs. However, the last three limits are highly
nontrivial: comparing the anomaly polynomials on the two sides of the equation
requires us to analytically continue to the case of zero $\mathfrak{g}%
_{\mathrm{max}}$ gauge algebras. Remarkably, we find a perfect match between
the two sides after performing this analytic continuation. Note also that the
$-3$ NHC is special in that it requires a homomorphism $\Gamma_{E_{6}}
\rightarrow E_{8}$. In the rest of the examples, we have chosen the convention of measuring the length of the quiver by $\ell_{\text{end}}$, but we could just have easily used the $\ell_{\text{plat}}$ convention.

Although the match in anomaly polynomials serves as the primary confirmation
of these limits, another cross-check comes from comparing the global symmetries of
the theories. In particular, the global symmetry of a limit theory always
contains the global symmetry of the theories in the large $\ell_{\text{plat}}$ limit.
For instance, all of the above NHCs have trivial global symmetry, as do the
quivers on the right-hand side of (\ref{eq:firstNHC})--(\ref{eq:lastNHC}). As
another example, the theory of a stack of M5-branes probing a $D_{4}$
singularity takes the form
\begin{equation}
\lbrack\mathfrak{so}(8)]\,\,1\,\,\overset{\mathfrak{so}(8)}{4}%
\,\,1\,\,\overset{\mathfrak{so}(8)}{4}\,\,1\,\,\cdots
\,\,\overset{\mathfrak{so}(8)}{4}\,\,1\,\,\overset{\mathfrak{so}%
(8)}{4}\,\,1\,\,[\mathfrak{so}(8)]
\end{equation}
which has $\mathfrak{so}(8)\oplus\mathfrak{so}(8)$ global symmetry. The
limiting case of a single M5-brane gives the rank 1 E-string theory,
\begin{equation}
\lbrack\mathfrak{e}_{8}]\,\,1
\end{equation}
Here, $\mathfrak{so}(8)\oplus\mathfrak{so}(8)\subset\mathfrak{e}_{8}$, so
indeed the global symmetry of the theory in the small $\ell_{\text{plat}}$
limit contains the global symmetry of the theory in the large $\ell
_{\text{plat}}$ limit. Similarly, the theory in (\ref{eq:G2SO13ex}) has
$\mathfrak{g}_{2}\oplus\mathfrak{so}(13)$ global symmetry, while the theories
in (\ref{eq:G2SO12ex}) have a strictly smaller $\mathfrak{g}_{2}%
\oplus\mathfrak{so}(12)$ global symmetry.

Similar considerations also hold even when the endpoint is trivial. For example,
in some cases, one must analytically continue all the way to a \emph{negative}
number of $\mathfrak{g}_{\mathrm{max}}$ gauge algebras. For instance, we may
write
\begin{equation}
\overset{\mathfrak{sp}(2)}{1}\,\,\overset{\mathfrak{g}_{2}}{2}%
\,\,\overset{\mathfrak{su}(2)}{2}=\lim_{\ell_{\text{plat}}\rightarrow
-1}\overset{\mathfrak{sp}(2)}{1}\,\,\overset{\mathfrak{so}(7)}{3}%
\,\,1\,\,\overset{\mathfrak{so}(8)}{4}\,\,\cdots
\,\,1\,\,\overset{\mathfrak{so}(8)}{4}\,\,1\,\,\underset{[\mathfrak{su}%
(2)]}{\overset{\mathfrak{so}(7)}{3}}\,\,\overset{\mathfrak{su}(2)}{2}%
\end{equation}
This match requires an analytic continuation to $\ell_{\text{plat}} = -1$ in the
number of $\mathfrak{so}%
(8)=\mathfrak{g}_{\mathrm{max}}$ gauge algebras. Once again, we stress that this continuation to a negative number of gauge algebras is merely a formal, mathematical operation. Note also that, unlike in the
previous examples, the $\mathfrak{g}_{2}$ gauge algebra that
appears in the theory on the left-hand side does not appear in the infinite
family of gauge algebras. Instead, the two $\mathfrak{so}(7)$ gauge algebras
separated by the chain of $\mathfrak{so}(8)$s have merged, in a sense, to
become a $\mathfrak{g}_{2}$ gauge algebra. Morally, we have $\mathfrak{so}%
(7)+\mathfrak{so}(7)-\mathfrak{so}(8)=\mathfrak{g}_{2}$. The appearance of a formal subtraction operation
suggests a corresponding role for addition and subtraction of 7-branes in F-theory,
as occurs for example in K-theory (namely formal addition and subtraction of vector bundles). This
would generalize the K-theoretic considerations for D-branes found in reference
\cite{Witten:1998cd} to F-theory.

\subsection{An Alternative Classification Scheme}

To what extent can the above approaches be considered a complete
classification? Using the classification of ``long bases" in appendix B of
\cite{Heckman:2015bfa}, it is a straightforward exercise to show that all
sufficiently long 6D SCFTs with A-type endpoints listed in (\ref{eq:Anendpoint}) can
be classified uniquely by the gauge algebra $\mathfrak{g}_{\mathrm{max}}$ as
well as a pair of nilpotent orbits $\mathcal{O}_{\mathrm{L}}$, $\mathcal{O}_{\mathrm{R}}$ of Lie
algebras $\mathfrak{g}_{\mathrm{L}}$, $\mathfrak{g}_{\mathrm{R}}$, where these Lie algebras are
the maximal flavor symmetry on the left and right of the quiver, respectively,
for the given endpoint and given choice of $\mathfrak{g}_{\mathrm{max}}$. Similarly, all sufficiently long
6D SCFTs with trivial endpoints can be uniquely classified by the gauge algebra $\mathfrak{g}_{\mathrm{max}}$, a discrete homomorphism $\Gamma \rightarrow E_8$, and a nilpotent orbit $\mathcal{O}_{\mathrm{R}}$ of the Lie algebra $\mathfrak{g}_{\mathrm{R}}$.

However, we will show in section \ref{sec:DARTHFUSIUS} that there seem to
be outliers that do not fit into the above schemes for $\ell_{\text{end}}%
\leq10$, $\ell_{\text{plat}}\leq8$. When $\ell_{\text{plat}}\geq9$, however,
every known 6D SCFT can be given a \emph{unique} description.
Measuring the size of a theory by $\ell_{\text{plat}}$, we see that the
overwhelming majority of 6D SCFTs can be classified using group theory.

Even for theories with long A-type endpoints $\ell_{\text{end}%
}\geq11$, there are subtleties. This occurs primarily when the rank
of the flavor symmetry algebra also becomes large, and is comparable to $\ell_{\text{end}}$,
as can happen in the case of classical flavor symmetry algebras of $\mathfrak{su}$, $\mathfrak{so}$ and $\mathfrak{sp}$ type.
For example, while every 6D SCFT with a long A-type endpoint
admits a group-theoretic description, this choice is not necessarily unique,
and some pairs of nilpotent orbits might not be allowed for a given endpoint.
This happens when the endpoint is too short relative to the size of the breaking pattern for
the flavor symmetries. In such cases the nilpotent deformation on
the left-hand side of the quiver can overlap with the nilpotent deformation on the
right-hand side of the quiver. For example, this can occur for nilpotent orbits of
$\mathfrak{su}(m)$ when $m$ becomes comparable to $\ell_{end}$, and similar considerations apply for the
$\mathfrak{so}$ and $\mathfrak{sp}$ cases as well. Note, however,
that even in this case, the analytic continuation of certain generic patterns
allows us to also cover a number of ``short'' (relative to the size of the nilpotent orbits)
bases in the same sort of classification scheme. This again indicates that nearly all 6D\ SCFTs can be
labeled in terms of simple algebraic data.

There are some outliers that still resist inclusion in this sort
of classification scheme. As we now show, however, even these cases are closely connected to
the fission products of our orbi-instanton progenitor theories.

\section{All Remaining Outliers from one Step of Fusion \label{sec:DARTHFUSIUS}}

It is rather striking that the vast majority of 6D\ SCFTs all descend from a
handful of progenitor orbi-instanton theories. As we have already remarked,
even theories with a low dimension tensor branch can often be viewed as
limiting cases. But there are also some outliers which do not fit
into such a taxonomy. Rather, such theories should better be viewed as
another class of progenitor theories.
This includes models with a small number of curves, as well as models
with a D- or E-type endpoint.

In this section we show that aside from the ADE\ $(2,0)$ theories, all of
these outliers are obtained through the process of fusion, in which we take possibly multiple decoupled 6D\ SCFTs and gauge a common
non-abelian flavor symmetry, pairing it with an additional tensor multiplet,
and move to the origin of the new tensor branch moduli space. The $(2,0)$ theories can all be reached by performing a
Higgs branch deformation of $(1,0)$ theories with the same endpoint configuration of $-2$ curves.

In some sense, the gluing rules for NHCs given for general F-theory backgrounds in reference \cite{Morrison:2012np} and
developed specifically for 6D SCFTs in reference \cite{Heckman:2013pva} already tell us that since all NHCs are fission products,
we can generate all 6D SCFTs via fusion. The much more non-trivial
feature of the present analysis is that after \textit{just one step} of fusion \textit{all} outlier 6D SCFTs (that is, those
theories not obtained from fission of the orbi-instanton theories) are realized.
The main reason to expect that something like this is possible is to observe that in nearly all configurations, a curve typically intersects
at most two other curves, and rarely intersects three or more. In those cases where a curve intersects three or more curves, it necessarily has a
gauge group attached to it, and this is the candidate ``fusion point'' which after blowup, takes us to a list of decoupled fission products
obtained from our progenitor orbi-instanton theories. Indeed, the appearance of such curves is severely restricted in 6D SCFTs,
and this is the main reason we should expect a single fusion step to realize all of our outlier theories.

In this section we first establish that there are indeed theories
which cannot be obtained from fission of the orbi-instanton
progenitor theories. After this, we show that all of these examples (as well
as many more)\ can be obtained through a simple fusion operation. While
we have not performed an exhaustive sweep over every 6D\ SCFT from the
classification of reference \cite{Heckman:2015bfa}, we already see that outlier
theories exhibit some structure, and within the corresponding patterns, we find no counterexamples
to the claim that all 6D\ SCFTs are products of either a single fission
operation or fission and then a further fusion operation.

\subsection{Examples of Outlier Theories}

We now turn to some examples of outlier theories. The reason all of these
examples cannot be obtained from a progenitor orbi-instanton theory has to do
with the structure of the anomaly polynomial for the orbi-instanton theories,
and the resulting tensor branch / homplex deformations. As obtained in
\cite{Mekareeya:2016yal}, all the descendant anomaly polynomials exhibit a clear
pattern which also persists upon \textquotedblleft analytic
continuation\textquotedblright\ in the parameters $\ell_{\text{plat}}$ and  $\ell_{\text{end}}$.
While that paper focused on the case of nilpotent orbit deformations, the statement also holds for the case of discrete homomorphisms $\Gamma \rightarrow E_8$. Thus, for a given $\gmax$, we can select two homplex deformations--one on the left and one on the right--and compute the anomaly polynomial of the resulting family of theories as a function of $\ell_{\text{plat}}$. For instance, for $\gmax = \mf{e}_8$, focusing on the theories with trivial endpoint, there are 70 nilpotent orbits of $\mf{e}_8$ and 137 homomorphisms $\Gamma_{E_8} \rightarrow E_8$ \cite{TRFrey}. This means there are $70 \times 137 = 9590$ families of theories to consider, each of which is parametrized by $\ell_{\text{plat}}$. If we instead focus on theories labeled by a pair of $\mf{e}_8$ nilpotent orbits, which necessarily have endpoint $22...2$, we have (by left-right symmetry) $70 \times 71 /2 = 2485$ distinct families of theories parametrized by $\ell_{\text{plat}}$. Using a computer sweep, we have computed the anomaly polynomials for all families with $\gmax =\mf{so}(8)$, $\mf{so}(10)$, $\mf{so}(12)$, $\mf{e}_6$, $\mf{e}_7$, and $\mf{e}_8$ as a function of $\ell_{\text{plat}}$. We find that there exist apparently consistent 6D SCFTs whose anomaly polynomials do not appear in any of the families indexed by $\ell_{\text{plat}}$, even allowing for analytic continuation to $\ell_{\text{plat}} \leq 0$.

Let us illustrate with explicit examples of such SCFTs. Consider, for instance, gauging
the $E_8$ flavor symmetry of the rank-10 E-string theory:
\begin{equation}\label{eq:rank10Estring}
\overset{\mathfrak{e}_{8}}{(12)}%
\,\,1\,\,2\,\,2\,\,2\,\,2\,\,2\,\,2\,\,2\,\,2\,\,2
\end{equation}
This has endpoint $2$ and $\mathfrak{g}_{\mathrm{max}}=\mathfrak{e}_{8}$, but
there is no way to generate via fission starting from a progenitor orbi-instanton theory.
This becomes especially clear if we add a ramp of $\mathfrak{su}(m_{i})$ gauge algebras:
\begin{equation}
\overset{\mathfrak{e}_{8}}{(12)}\,\,1\,\,2\,\,\overset{\mathfrak{su}%
(2)}{2}\,\,\overset{\mathfrak{su}(3)}{2}\,\,\overset{\mathfrak{su}%
(4)}{2}\,\,\overset{\mathfrak{su}(5)}{2}\,\,\overset{\mathfrak{su}%
(6)}{2}\,\,\overset{\mathfrak{su}(7)}{2}\,\,\overset{\mathfrak{su}%
(8)}{2}\,\,\overset{\mathfrak{su}(9)}{2}\,\,[\mathfrak{su}(10)]
\end{equation}
The flavor symmetry here is $\mathfrak{su}(10)$, which is not a subalgebra of
$\mathfrak{e}_{8}$, hence there is no way to realize this as the commutant of
a nilpotent orbit of $\mathfrak{e}_{8}$. The fact that this outlier does not
fit into our previous classification is related to the fact that there is no
analogous class $\mathcal{S}$ theory in 4D, as discussed on page 35 of
\cite{Ohmori:2015pia}.

Some examples can be viewed as the collisions of different
singularities. For such examples, there is some hope that these particular
outliers could be labeled by group theoretic data. In particular, we can view
such an outlier as a collision of two homomorphisms. As a simple example, we
start with the $\ell_{\text{end}}=1$ $\mathfrak{e}_{8}$ theory with
$\mathcal{O}_{\mathrm{L}}=\mathcal{O}_{\mathrm{R}}=A_{4}+A_{3}$:
\begin{equation}
2\,\,2\,\,2\,\,2\,\,1\,\,\overset{\mathfrak{e}_{8}}{(12)}%
\,\,1\,\,2\,\,2\,\,2\,\,2
\end{equation}
If we blow down the small instanton chains on the left and right of the $-12$
curve, we are left with a $-2$ curve carrying $\mathfrak{e}_{8}$ gauge algebra
with two singular marked points indicating the location where the small
instantons were blown down. If we collide these two marked points, the model
becomes even more singular, and the resolution produces the theory in
(\ref{eq:rank10Estring}).

As a more nontrivial example, we consider the $\mathfrak{g}_{\mathrm{max}%
}=\mathfrak{e}_{8}$, $\ell_{\text{plat}}=1$ theory with $\mathcal{O}_{\mathrm{R}}$ the
$A_{4}$ nilpotent orbit of $\mathfrak{e}_{8}$ and $\mathcal{O}_{\mathrm{L}}$ the
$A_{2}$ orbit of $\mathfrak{f}_{4}$:
\begin{equation}\label{eq:twonilpotent}
\overset{\mathfrak{su}(1)}{2}\,\,\overset{\mathfrak{su}(2)}{2}%
\,\,\overset{\mathfrak{su}(1)}{2}\,\,1\,\,\overset{\mathfrak{e}_{8}%
}{(12)}\,\,1\,\,\overset{\mathfrak{su}(1)}{2}\,\,\overset{\mathfrak{su}%
(2)}{2}\,\,\overset{\mathfrak{su}(3)}{2}\,\,\overset{\mathfrak{su}%
(4)}{2}\,\,[\mathfrak{su}(5)]
\end{equation}
Blowing down the small instanton chains on the left and right, we get a $-3$
curve supporting a $II^{\ast}$ fiber, and further singularities
at marked points of this curve. Adopting local coordinates $(s,t)$, this
is described by a Weierstrass model of the form $y^2 = x^3 + fx + g$ with $\Delta = 4 f^3 + 27 g^2$, where we have the
local presentation:
\begin{align}
f  &  =-\frac{1}{48}s^{4}+\frac{1}{6}s^{4}t^{2}(\epsilon-t)^{4}-\frac{1}%
{6}s^{4}t(\epsilon-t)^{2}+s^{5}t^{5}\label{eq:f}\\
g  &  =\frac{1}{864}s^{6}+\frac{1}{4}s^{6}t^{4}(\epsilon-t)^{8}-\frac{5}%
{54}s^{6}t^{3}(\epsilon-t)^{6}+\frac{1}{72}s^{6}t^{2}(\epsilon-t)^{4}+\frac
{1}{72}s^{6}t(\epsilon-t)^{2}+s^{5}t^{5}(\epsilon-t)^{4}\\
\Delta & = s^{10} t^5 \left(   \frac{s^3}{192}  + 27 (\epsilon - t)^8 t^5 + \cdots \right).
\end{align}
Here, the $-3$ curve is given by $\{s=0\}$, the orbit $\mathcal{O}_{\mathrm{L}}$
corresponds to the point $s=0,t=\epsilon$, and the $\mathcal{O}_{\mathrm{R}}$ orbit
corresponds to the point $s=0,t=0$. Colliding the two singularities then
corresponds to taking $\epsilon\rightarrow0$. Resolving the singular point
$t=0$, we get an outlier theory of the form
\begin{equation}
\overset{\mathfrak{e}_{8}}{(12)}\,\,1\,\,\overset{\mathfrak{su}(1)}{2}%
\,\,\overset{\mathfrak{su}(2)}{2}\,\,\overset{\mathfrak{su}(3)}{2}%
\,\,\overset{\mathfrak{su}(4)}{2}\,\,\overset{\mathfrak{su}(5)}{2}%
\,\,\overset{\mathfrak{su}(6)}{2}\,\,\underset{[\mathfrak{su}%
(2)]}{\overset{\mathfrak{su}(7)}{2}}\,\,\overset{\mathfrak{su}(6)}{2}%
\,\,[\mathfrak{su}(5)] \label{eq:oneoutlier}%
\end{equation}
We might have expected this result from looking at the original theory in
(\ref{eq:twonilpotent}): the drop-off in gauge algebra ranks on the left-hand
side of (\ref{eq:twonilpotent}) has been superimposed on the ramp of gauge
algebra ranks on the right-hand side of (\ref{eq:twonilpotent}).

As another example, if we begin with a 6D SCFT with tensor branch:
\begin{equation}
\overset{\mathfrak{su}(2)}{2}\,\,\overset{\mathfrak{su}(2)}{2}%
\,\,\overset{\mathfrak{su}(2)}{2}\,\,\overset{\mathfrak{su}(1)}{2}%
\,\,1\,\,\overset{\mathfrak{e}_{8}}{(12)}\,\,1\,\,\overset{\mathfrak{su}%
(1)}{2}\,\,\overset{\mathfrak{su}(2)}{2}\,\,\overset{\mathfrak{su}%
(3)}{2}\,\,\overset{\mathfrak{su}(4)}{2}\,\,[\mathfrak{su}(5)]
\end{equation}
and once again blow down the exceptional divisors, collide the singular
points, and resolve, we expect the resulting theory to be:
\begin{equation}
\overset{\mathfrak{e}_{8}}{(12)}\,\,1\,\,\overset{\mathfrak{su}(1)}{2}%
\,\,\overset{\mathfrak{su}(2)}{2}\,\,\overset{\mathfrak{su}(3)}{2}%
\,\,\overset{\mathfrak{su}(4)}{2}\,\,\overset{\mathfrak{su}(5)}{2}%
\,\,\overset{\mathfrak{su}(6)}{2}\,\,\underset{[N_{f}%
=1]}{\overset{\mathfrak{su}(7)}{2}}\,\,\overset{\mathfrak{su}(7)}{2}%
\,\,\overset{\mathfrak{su}(7)}{2}\,\,[\mathfrak{su}(7)]
\end{equation}
This can indeed be achieved via the Weierstrass model over a $-2$ curve with a $II^{\ast}$ fiber
and further singularities at marked points of the curve. Again using local coordinates $(s,t)$ with the
$-2$ curve at $s = 0$, the corresponding $(f,g,\Delta)$ are:
\begin{align}
f  &  =-\frac{1}{48}s^{4}+s^{4}t^{5}(\epsilon -t)^{2}+\frac{1}{6}s^{4}t^{2}%
(\epsilon -t)^{4}-\frac{1}{6}s^{4}z(\epsilon - t)^{2}\\
g  &  =\frac{1}{864}s^{6}+\frac{1}{4}s^{6}t^{4}(\epsilon -t)^{8}-\frac{5}{54}%
s^{6}t^{3}(\epsilon - t)^{6}+\frac{1}{72}s^{6}t^{2}(\epsilon - t)^{4}+\frac{1}{72}%
s^{6}t(\epsilon - t)^{2}+2s^{5}t^{5}(\epsilon - t)^{5}\\
\Delta & = s^{10} t^5 (\epsilon-t)^2 \left(  \frac{s^2}{192} + 108 (\epsilon - t)^8 t^5 + \cdots \right).
\end{align}

Outliers can be more complicated, however; they do not always seem to arise in
this manner of colliding singularities. Consider, for instance, a chain of
four $\mathfrak{e}_{6}$ gauge algebras, and add a small instanton to the
second one:
\begin{equation}
\overset{\mathfrak{e}_{6}}{6}\,\,1\,\,\overset{\mathfrak{su}(3)}{3}%
\,\,1\,\,\underset{1}{\overset{\mathfrak{e}_{6}}{6}}%
\,\,1\,\,\overset{\mathfrak{su}(3)}{3}\,\,1\,\,\overset{\mathfrak{e}_{6}%
}{6}\,\,1\,\,\overset{\mathfrak{su}(3)}{3}\,\,1\,\,\overset{\mathfrak{e}%
_{6}}{6} \label{eq:E6outlier}%
\end{equation}
Performing two iterations of blowing down the $-1$ curves, we find a
configuration of the form:
\begin{equation}
4124
\end{equation}
We see why this theory cannot be realized as the limit of an infinite family:
this would require a large number of $\mathfrak{e}_{6}$ gauge algebras, which
would blow down to
\begin{equation}
41\underset{\ell_{2}}{\underbrace{22...22}}4.
\end{equation}
But such configurations are only valid when $\ell_{2} < 3$; at larger values
the intersection pairing ceases to be negative definite. We have checked that this
theory is indeed an outlier by computing the anomaly polynomials for
\emph{all} theories labeled by a pair of homomorphisms with $\mathfrak{g}%
_{\mathrm{max}}=\mathfrak{so}(2 m),\mathfrak{e}_{n}$, using the theories
computed in \cite{TRFrey}. For a given pair of homomorphisms and a given
$\mathfrak{g}_{\mathrm{max}}$, we analytically continue to $\ell
_{\text{plat}}\leq0$. The anomaly polynomial for the theory in
(\ref{eq:E6outlier}) never appears in the list of analytic continuations,
showing that this theory truly is an \textquotedblleft outlier": it cannot be
produced by the process of \textquotedblleft fission" described previously.

This is an example of a larger set of outliers, which blow down to
some configuration of the form
\begin{equation}
\alpha12...2\beta\label{eq:outlierset}%
\end{equation}
with $\alpha, \beta$ representing some configurations of curves of
self-intersection $-2$ or below. For instance, we may have
\begin{equation}
\overset{\mathfrak{su}(3)}{3 }\,\, 1 \,\, \overset{\mathfrak{e}_{6}}{6 }\,\, 1
\,\, \overset{\mathfrak{su}(3)}{3 }\,\, 1 \,\, {\overset{\mathfrak{f}_{4}}{5}}
\,\,1 \,\, \overset{\mathfrak{su}(3)}{3 }\,\, 1 \,\, \overset{\mathfrak{e}%
_{6}}{6 }\,\,1 \,\, \overset{\mathfrak{su}(3)}{3 }\,\, 1 \,\,
\overset{\mathfrak{e}_{6}}{6}%
\end{equation}
This blows down to
\begin{equation}
2 3 1 2 4
\end{equation}
Similarly, we can consider
\begin{align}
{\overset{\mathfrak{e}_{7}}{8}} \,\, 1 \,\, \overset{\mathfrak{su}(2)}{2 }\,\,
\overset{\mathfrak{so}(7)}{3 }\,\, \overset{\mathfrak{su}(2)}{2 }\,\, 1 \,\,
\overset{\mathfrak{e}_{7}}{7 }\,\, 1 \,\, \overset{\mathfrak{su}(2)}{2 }\,\,
\overset{\mathfrak{so}(7)}{3 }\,\, \overset{\mathfrak{su}(2)}{2 }\,\, 1 \,\,
\overset{\mathfrak{e}_{7}}{8 }\,\, 1 \,\, \overset{\mathfrak{su}(2)}{2 }\,\,
\overset{\mathfrak{so}(7)}{3 }\,\, \overset{\mathfrak{su}(2)}{2 }\,\, 1 \,\,
\overset{\mathfrak{e}_{7}}{8 }\,\, 1 \,\, \overset{\mathfrak{su}(2)}{2 }\,\,
\overset{\mathfrak{so}(7)}{3 }\,\, \overset{\mathfrak{su}(2)}{2 }\,\, 1 \,\,
\overset{\mathfrak{e}_{7}}{8 }\,\, 1 \,\, \overset{\mathfrak{su}(2)}{2 }\,\,
\overset{\mathfrak{g}_{2}}{3} \label{eq:E7outlier}%
\end{align}
This blows down to
\begin{equation}
512232
\end{equation}

In section \ref{sec:DARTHFISSIUS}, we claimed that our classification gives the complete set of 6D SCFTs for $\ell_{\text{end}}\geq11$,
$\ell_{\text{plat}}\geq9$. We now see where these numbers come from: consider
the set of outliers in (\ref{eq:outlierset}) with $\mathfrak{g}_{\mathrm{max}%
}=\mathfrak{e}_{8}$, and take $\alpha=\beta=22223$. This blows down to an
outlier with endpoint $2222222222$, which has $\ell_{\text{end}}=10$.
Similarly, taking $\alpha=\beta=7$ gives an outlier with $\ell_{\text{plat}%
}=8$ $\mathfrak{e}_{8}$ gauge algebras. These appear to be the largest
outliers, so for $\ell_{\text{end}}$ and $\ell_{\text{plat}}$ above this,
every theory fits into our classification strategy.

\subsection{Fusion Products}

All of the outlier theories we have seen so far fall under the same pattern:
we take one or more \textquotedblleft fission" 6D SCFT and gauge a common
flavor symmetry. In the case of (\ref{eq:oneoutlier}), for instance, we gauge
an $\mathfrak{e}_{8}$ flavor symmetry:
\begin{equation}
\lbrack\mathfrak{e}_{8}]\,\,1\,\,\overset{\mathfrak{su}(1)}{2}%
\,\,\overset{\mathfrak{su}(2)}{2}\,\,\overset{\mathfrak{su}(3)}{2}%
\,\,\overset{\mathfrak{su}(4)}{2}\,\,\overset{\mathfrak{su}(5)}{2}%
\,\,\overset{\mathfrak{su}(6)}{2}\,\,\underset{[\mathfrak{su}%
(2)]}{\overset{\mathfrak{su}(7)}{2}}\,\,\overset{\mathfrak{su}(6)}{2}%
\,\,[\mathfrak{su}(5)]
\end{equation}
In (\ref{eq:E6outlier}), we gauge the common $\mathfrak{e}_{6}$ flavor
symmetry of three fission products to produce a single fusion product:
\begin{align}
\overset{\mathfrak{e}_{6}}{6}\,\,1\,\,\overset{\mathfrak{su}(3)}{3}%
\,\,1\,\,[\mathfrak{e}_{6}]  &  ~\oplus~[\mathfrak{e}_{6}%
]\,\,1\,\,\overset{\mathfrak{su}(3)}{3}\,\,1\,\,\overset{\mathfrak{e}_{6}%
}{6}\,\,1\,\,\overset{\mathfrak{su}(3)}{3}\,\,1\,\,\overset{\mathfrak{e}%
_{6}}{6}\nonumber\\
&  \overset{[\mathfrak{e}_{6}]}{\underset{[\mathfrak{su}(3)]}{1}}%
\end{align}

Typically, a given theory may result from fusion in multiple ways. To get the
theory of line (\ref{eq:E7outlier}), we can gauge the leftmost $\mathfrak{e}_{7}$ flavor
symmetry of the fission:
\begin{equation}
[\mathfrak{e}_{7}] \,\, 1 \,\, \overset{\mathfrak{su}(2)}{2 }\,\,
\overset{\mathfrak{so}(7)}{3 }\,\, \overset{\mathfrak{su}(2)}{2 }\,\, 1 \,\,
\overset{\mathfrak{e}_{7}}{7 }\,\, 1 \,\, \overset{\mathfrak{su}(2)}{2 }\,\,
\overset{\mathfrak{so}(7)}{3 }\,\, \overset{\mathfrak{su}(2)}{2 }\,\, 1 \,\,
\overset{\mathfrak{e}_{7}}{8 }\,\, 1 \,\, \overset{\mathfrak{su}(2)}{2 }\,\,
\overset{\mathfrak{so}(7)}{3 }\,\, \overset{\mathfrak{su}(2)}{2 }\,\, 1 \,\,
\overset{\mathfrak{e}_{7}}{8 }\,\, 1 \,\, \overset{\mathfrak{su}(2)}{2 }\,\,
\overset{\mathfrak{so}(7)}{3 }\,\, \overset{\mathfrak{su}(2)}{2 }\,\, 1 \,\,
\overset{\mathfrak{e}_{7}}{8 }\,\, 1 \,\, \overset{\mathfrak{su}(2)}{2 }\,\,
\overset{\mathfrak{g}_{2}}{3}%
\end{equation}
or we can gauge the common flavor symmetry of the two fission products:
\begin{equation}
{\overset{\mathfrak{e}_{7}}{8}} \,\, 1 \,\, \overset{\mathfrak{su}(2)}{2 }\,\,
\overset{\mathfrak{so}(7)}{3 }\,\, \overset{\mathfrak{su}(2)}{2 }\,\, 1 \,\,
[\mathfrak{e}_{7}] ~\oplus~[\mathfrak{e}_{7}] \,\, 1 \,\,
\overset{\mathfrak{su}(2)}{2 }\,\, \overset{\mathfrak{so}(7)}{3 }\,\,
\overset{\mathfrak{su}(2)}{2 }\,\, 1 \,\, \overset{\mathfrak{e}_{7}}{8 }\,\, 1
\,\, \overset{\mathfrak{su}(2)}{2 }\,\, \overset{\mathfrak{so}(7)}{3 }\,\,
\overset{\mathfrak{su}(2)}{2 }\,\, 1 \,\, \overset{\mathfrak{e}_{7}}{8 }\,\, 1
\,\, \overset{\mathfrak{su}(2)}{2 }\,\, \overset{\mathfrak{so}(7)}{3 }\,\,
\overset{\mathfrak{su}(2)}{2 }\,\, 1 \,\, \overset{\mathfrak{e}_{7}}{8 }\,\, 1
\,\, \overset{\mathfrak{su}(2)}{2 }\,\, \overset{\mathfrak{g}_{2}}{3}%
\end{equation}

This illustrates that for theories with A-type (and trivial) endpoints,
various outlier theories can all be realized via fusion operations.
Let us now turn to theories with D- and E-type endpoints.

\subsubsection{D- and E-type Endpoints}

6D SCFTs with D- and E-type endpoints are distinguished by the presence of a
trivalent node that remains even after blowing down all $-1$ curves of the
base. We now show that (excluding the (2,0) SCFTs), these theories can also be
produced by the aforementioned process of ``fusion," where we gauge the flavor
symmetry associated with the trivalent node.

As in the A-type case, we begin by looking at long endpoints i.e.~those with
large $\ell_{\text{end}}$. At sufficiently large $\ell_{\text{end}}$, the possible 6D SCFTs are very limited.
Clearly, there are no E-type endpoints for large $\ell_{\text{end}}$. For D-type endpoints
with $\mathfrak{g}_{\mathrm{max}} = \mathfrak{su}(2m)$, we have
\begin{equation}
\overset{\mathfrak{su}(m)}{2 }\,\, \overset{\mathfrak{su}%
(m)}{\overset{2}{\overset{\mathfrak{su}(2m)}{2}}} \,\,\overset{\mathfrak{su}%
(2m)}{2 }\,\,\cdots\,\,\overset{\mathfrak{su}(m_{\ell_{\text{end}}-3})}{2 }%
\,\,\overset{\mathfrak{su}(m_{\ell_{\text{end}} - 2})}{2} \label{eq:longDtype}%
\end{equation}
Here, the possible choices of $m_{i}$ at the right-hand side of the quiver are
determined by partitions of $2m$, as in the A-type case. The choices of
$m_{i}$ at the left-hand side of the quiver are uniquely fixed for large $\ell_{\text{end}}$. The only other
option is $\mathfrak{g}_{\mathrm{max}} = \mathfrak{e}_{6}$. In this case, we
have a tensor branch description:
\begin{equation}
\overset{\mathfrak{su}(3)}{3 }\,\,1\,\, \overset{\mathfrak{su}%
(3)}{\overset{3}{\overset{1}{\overset{\mathfrak{e}_{6}}{6}}}} \,\, 1 \,\,
\overset{\mathfrak{su}(3)}{3 }\,\,1 \,\, \cdots
\end{equation}
The left-hand side is fixed, while the right-hand side is
simply the right-hand side of an A-type quiver with $\ell_{\text{end}}$ large, $\mathfrak{g}%
_{\mathrm{max}} = \mathfrak{e}_{6}$, so it is labeled in terms of an
appropriate nilpotent orbit.

At small $\ell_{\text{end}}$, there are theories that do not fit into the above families of 6D
SCFTs. For instance, for the endpoint of type $D_{4}$:
\begin{equation}
2 \overset{2}{2}2
\end{equation}
we have a variety of theories such as:
\begin{equation}
\overset{\mathfrak{g}_{2}}{3 }\,\, \overset{\mathfrak{su}(2)}{2 }\,\, 1 \,\,
\overset{\mathfrak{g}_{2}}{\overset{3}{\overset{\mathfrak{su}%
(2)}{\overset{2}{\overset{1}{\overset{\mathfrak{e}_{8}}{8}}}}}} \,\, 1
\,\,\ \overset{\mathfrak{su}(2)}{2}\,\, \overset{\mathfrak{g}_{2}}{3}
\label{eq:shortDtype}%
\end{equation}

In all cases, we can realize these theories as fusions of three fission products
by gauging the common flavor symmetry. For (\ref{eq:longDtype}), we gauge the
common $\mathfrak{su}(2m)$ of:
\begin{align}
&  \underset{[\mathfrak{su}(2m)]}{\overset{\mathfrak{su}(m)}{2}}\nonumber\\
\overset{\mathfrak{su}(m)}{2 }\,\, [\mathfrak{su}(2m)]  &  ~\oplus~
[\mathfrak{su}(2m)] \,\,\overset{\mathfrak{su}(2m)}{2 }\,\,\cdots
\,\,\overset{\mathfrak{su}(m_{\ell_{\text{end}} - 3})}{2 }\,\,\overset{\mathfrak{su}(m_{\ell_{\text{end}} - 2})}{2}%
\end{align}
For (\ref{eq:shortDtype}), we gauge the common $\mathfrak{e}_{7}$ of:
\begin{align}
&  \underset{[\mathfrak{e}_{7}]}{\overset{\mathfrak{g}_{2}%
}{\overset{3}{\overset{\mathfrak{su}(2)}{\overset{2}{1}}}}}\nonumber\\
\overset{\mathfrak{g}_{2}}{3 }\,\, \overset{\mathfrak{su}(2)}{2 }\,\, 1 \,\,
[\mathfrak{e}_{7}]  &  ~\oplus~ [\mathfrak{e}_{7}] \,\, 1
\,\,\ \overset{\mathfrak{su}(2)}{2}\,\, \overset{\mathfrak{g}_{2}}{3}%
\end{align}
This method of fusion does not work for the $(2,0)$ theories of D- and E-type,
since these theories do not have gauge symmetries. Nevertheless, they can be reached
rather easily from a Higgs branch flow involving a $(1,0)$ theory with the same endpoint.
Neglecting this trivial case, all $(1,0)$ SCFTs with D- and E-type endpoints are fusion products.

Putting all of the pieces together, we see overwhelming evidence that all 6D SCFTs are
generated by at most one step of fission and one step of fusion.

\section{Towards the Classification of 6D RG Flows \label{sec:FLOWS}}

In this section we explain how the results of the previous section point the
way towards a classification of 6D\ RG\ flows. Starting from a UV\ fixed point,
suppose there is a supersymmetric deformation which generates a flow
from a UV fixed point to an IR fixed point. We claim there is then a flow where we first
perform a tensor branch deformation and then a Higgs branch flow.
The reason such a factorization is possible is simply because a complex
structure deformation cannot alter the singular base of a 6D\ SCFT, namely
$\mathbb{C}^{2}/\Gamma_{U(2)}$ with $\Gamma_{U(2)}$ a finite subgroup of
$U(2)$. For this reason, to perform a flow to a less singular model, we can
always first do the K\"{a}hler deformations of the base, and then hold fixed
the new singular base. Of course, there may be other trajectories in which we
alternate back and forth between tensor and Higgs branch flows.

In the previous sections, we showed that there is a remarkably streamlined
way to label 6D\ SCFTs, namely as the products of fission and fusion starting
from a small set of progenitor theories. Based on this simple algebraic
taxonomy, and given the fact that such algebraic structures also come with a
canonical partial ordering, this leads to a partial classification of
6D\ RG\ flows. The main idea here is that we look at nilpotent elements in the
semi-simple flavor symmetry $\mathfrak{g}_{\mathrm{L}}\times\mathfrak{g}_{\mathrm{R}}$ (rather
than just one factor), and order the orbits $\mathcal{O}^{\prime}%
\prec\mathcal{O}$ whenever $\mathcal{O}^{\prime}\subseteq\overline
{\mathcal{O}}$. This partial ordering on nilpotent orbits then corresponds to the ordering of theories under RG flows, with the UV fixed point labeled by the smaller orbit \cite{Heckman:2016ssk}.

While there is a known partial ordering for nilpotent orbits, it
is still evidently an open question as to whether there is a corresponding
partial order on homomorphisms $\Gamma_{ADE}\rightarrow E_{8}$, with $\Gamma_{ADE}$ a finite subgroup of
$SU(2)$.\footnote{We thank D. Frey and E. Witten for helpful discussions on this
point.} Turning the discussion around, we see that the natural partial ordering
suggested by complex structure deformations of an F-theory model, or equivalently Higgs branch flows of the 6D SCFT, actually motivates a partial ordering for the corresponding algebraic structure. Some aspects of this physically motivated
partial ordering were already noted in section \ref{sec:FISSFUSS}. For those discrete group homomorphisms which
define the same orbit in $E_8$ as a continuous homomorphism $\mathfrak{su}(2)\rightarrow \mathfrak{e}_8$ and its exponentiation to the Lie group,
we can simply borrow the partial ordering on nilpotent orbits for $\mathfrak{e}_8$. The complications arise from the fact that not
all discrete group homomorphisms lift in this way. An analysis of how to order these other cases is beyond the scope of the present work.

But in addition to these specialized Higgs branch deformations, there is also
a broader class of complex structure deformations that do not appear to
correspond to any homplex deformation. Rather, they are associated with
semi-simple elements of the flavor symmetry algebra, that is, deformations
valued in the Cartan subalgebra of a flavor symmetry.\footnote{Let us note that in the case of $U(1)$ flavor symmetries,
additional care is needed because a single $U(1)$ may act on flavors localized at different nodes
of the generalized quiver. In this case, identifying a common $U(1)$ really involves
specifying particular flavor fields.} The way this works in practice
is that we identify a common Cartan subalgebra for a pair of flavor symmetry factors,
and perform a brane recombination with respect to these factors. It is essentially a generalization of
brane recombination moves for conformal matter such as \cite{DelZotto:2014hpa, Heckman:2014qba}:
\begin{equation}
y^{2}=x^{3}+s^{5}t^{5}\rightarrow x^{3}+(st-r)^{5}.\label{recombo}
\end{equation}
Although these are straightforward to identify in the minimal
Weierstrass model, there is apparently no canonical partial ordering available just
from the structure of a Lie algebra.

Physically, however, there is a natural structure, as dictated by the way these semi-simple
deformations induce Higgs branch flows. The key difference between semi-simple deformations and nilpotent
deformations is that in a sufficiently long progenitor theory, a nilpotent deformation eventually terminates,
namely it leaves intact the interior algebra $\mathfrak{g}_{\text{max}}$ found on a partial tensor branch. By their very nature, we
see that in a semi-simple deformation, even satisfying the triplet of D-term constraints for the gauge groups
of the partial tensor branch of a progenitor theory correlates the vevs for conformal matter, so the deformation propagates across multiple
nodes of a quiver-like theory. In terms of the partial tensor branch description
given by a generalized quiver, we identify a common $\mathfrak{u}(1)^{r}$ subalgebra for a pair of flavor symmetry factors. Next, we draw
the minimal path on the quiver between these flavor symmetries. For each gauge group factor of the quiver which is part of the path,
the semi-simple deformation leads to a breaking pattern for the corresponding flavor group to
$\mathfrak{g}_{i}^{\prime} \times \mathfrak{u}(1)^{r} \subset \mathfrak{g}_{i}$, but leaves the other
gauge groups and flavor symmetry factors unchanged.

In general, there are two kinds of semi-simple deformations which are clearly different in the limit of a sufficiently long quiver.
First, there are those which are localized at one end of the quiver. This can
happen if a previous nilpotent deformation generates a flavor symmetry with at least two
flavor symmetry factors on one side of the quiver. For these cases, there is a clear connection to a
nilpotent deformation, and this fits with the fact that each nilpotent element $\mu \in \mathfrak{g}_{\text{flav}}$
canonically defines a generator of the Cartan subalgebra via $[\mu , \mu^\dag]$.

Second, there are those deformations which are not localized, namely they stretch from one side of the quiver to the other.
These cases cannot be associated with a nilpotent deformation (which eventually terminates in the interior of the quiver),
and so in this sense are genuine examples of semi-simple deformations.

Let us illustrate how this works in a few examples. Consider, for example,
the quiver:
\begin{equation}
\underset{[S(U(1)_{1}\times U(1)_{2})]}{\overset{\mathfrak{su}(2)}{2}%
,\overset{\mathfrak{su}(3)}{2}},\overset{\mathfrak{su}(3)}{2}%
,...,\overset{\mathfrak{su}(3)}{2},[SU(3)]
\end{equation}
which we obtain from a nilpotent deformation of the related quiver with all
$\mathfrak{su}(3)$ gauge algebra factors. The notation $[S(U(1)_1 \times U(1)_2)]$
reflects the fact that according to the computation of flavor symmetries performed in
\cite{Heckman:2016ssk}, the flavor symmetry actually has a single $U(1)$. This
flavor symmetry is delocalized in the sense that it is shared by the two $I_1$ loci which
intersect the partial tensor branch. For some additional discussion on $U(1)$ flavor
symmetries in F-theory which do not use the additional data provided from field theory,
see e.g. \cite{Bertolini:2015bwa, Lee:2018ihr}.

In this case, we can activate a semi-simple deformation by drawing a path between the two $U(1)$ flavor
symmetry factors, or by drawing a path between one of the $U(1)$ factors and
the right-hand side $SU(3)$ flavor symmetry factor. The resulting deformed
theory for these three cases is:
\begin{align}
  \text{Path}(U(1)_1,U(1)_2) : &  \overset{\mathfrak{su}(1)}{2}%
,\overset{\mathfrak{su}(2)}{2},\underset{[SU(1)]}{\overset{\mathfrak{su}%
(3)}{2}},...,\overset{\mathfrak{su}(3)}{2},[SU(3)]\\
  \text{Path}(U(1)_1,SU(3)) : & \overset{\mathfrak{su}(1)}{2}%
,\underset{[SU(1)]}{\overset{\mathfrak{su}(2)}{2}},\overset{\mathfrak{su}%
(2)}{2},...,\overset{\mathfrak{su}(2)}{2},[SU(2)]\\
  \text{Path}(U(1)_2,SU(3)) : & \,\,\, [SU(2)],\overset{\mathfrak{su}(2)}{2}%
,\overset{\mathfrak{su}(2)}{2},\overset{\mathfrak{su}(2)}{2}%
,...,\overset{\mathfrak{su}(2)}{2},[SU(2)]
\end{align}
Let us further note that the first possibility is localized on one side of the
quiver, and can be alternatively viewed as being generated from a nilpotent
deformation of the quiver with all $\mathfrak{su}(3)$ gauge algebras.

Consider next a generalized quiver with partial tensor branch description:
\begin{equation}
[\mathfrak{g}_0]-\mathfrak{g}_1-...-\mathfrak{g}_k - [\mathfrak{g}_{k+1}].
\end{equation}
We can activate a semi-simple deformation by tracing a path from the very left to the very right.
After the semi-simple deformation, this yields:
\begin{equation}\label{FuriousStyles}
[\mathfrak{g}^{\prime}_0]-\mathfrak{g}^{\prime}_1-...-\mathfrak{g}^{\prime}_k - [\mathfrak{g}^{\prime}_{k+1}],
\end{equation}
with $\mathfrak{g}_{i}^{\prime} \times \mathfrak{u}(1)^r \subset \mathfrak{g}_{i}$ for $i = 0,...,k+1$.

In all these cases, we see there is a clear notion of partial ordering for
semi-simple deformations based on containment relations for the
$\mathfrak{g}_{\text{max}}$ algebras which are left unbroken by the deformation.

Putting this together, we can also split up a complex structure
deformation for a generic quiver-like theory splits into two pieces:
\begin{itemize}
\item First, perform a semi-simple deformation.

\item Second, perform a homplex deformation.
\end{itemize}
This supplements the notion of fission developed here. Note that in the second
step, the homplex deformation is in some sense inherited from the parent
flavor symmetry.

Classifying which pairs of theories can be connected by Higgs branch flows then reduces to a mathematical question
of determining which nilpotent orbits of an algebra (after a semi-simple deformation) lift to a nilpotent orbit in
the parent algebra. If the lift of some nilpotent orbit $\mathcal{O}^\prime$ of $\mathfrak{g}^{\prime}_0$ is contained in the closure of an orbit $\mathcal{O}$ of $\mathfrak{g}_0$, there will be an RG flow involving a semi-simple deformation (and possibly a further homplex deformation) from the theory labeled by $\mathcal{O}$ to the theory labeled by $\mathcal{O}^\prime$.

\subsection{Examples}

It is helpful to illustrate the above considerations with some
explicit examples. To see how semi-simple deformations enter the
analysis of RG\ flows, consider the worldvolume theory of $k$ M5-branes
probing a $\mathbb{C}^{2}/\Gamma_{E_{8}}$ singularity. The partial tensor
branch for this theory is:%
\begin{equation}
\lbrack E_{8}]\,\,\underset{k - 1}{\underbrace{\overset{\mathfrak{e}_{8}%
}{2}\,\,\cdots\,\,\overset{\mathfrak{e}_{8}}{2}}}\,\,[E_{8}].
\end{equation}
Here, $k - 1$ is the number of $\mathfrak{e}_{8}$ gauge algebras in the quiver.
The F-theory model for the corresponding 6D\ SCFT can be described by the
intersection of 6D\ conformal matter with an $A_{k - 1}$ singularity:%
\begin{equation}
(y^{2}  = x^{3}+S^{5}T^{5}) \cap (ST  = W^{k}).
\end{equation}
We can also work in terms of covering space coordinates $(s,t)$ on
$\mathbb{C}^2$, namely we write:%
\begin{equation}
y^{2}  = x^{3}+s^{5}t^{5}\,\,\,\text{and}\,\,\, (s,t) \sim(\omega s,\omega^{-1}t)\,\,\,\text{with}\,\,\,\omega=\exp(2\pi i/k).
\end{equation}
The two coordinate systems are related via the $%
\mathbb{Z}
_{k}$ invariant combinations:%
\begin{equation}
s^{k}=S\text{, \ \ }t^{k}=T\text{, \ \ }st=W\text{.}%
\end{equation}
Now, we can perform a complex structure deformation by passing to the model:%
\begin{equation}
(y^{2}  = x^{3}+S^{5}T^{5}+\varepsilon(S^{3}T^{3})x) \cap (ST  = W^{k}),
\end{equation}
so we instead have a collision of two $E_{7}$ singularities at the $A_{k - 1}$
singularity. In terms of the M5-brane picture, the new partial tensor branch
is:%
\begin{equation}
\lbrack E_{7}]\,\,\underset{k - 1}{\underbrace{\overset{\mathfrak{e}_{7}%
}{2}\,\,\cdots\,\,\overset{\mathfrak{e}_{7}}{2}}}\,\,[E_{7}].
\end{equation}
We observe that for $k$ sufficiently large, there is no pair of nilpotent
orbits available which can produce a flow to this sort of theory. It seems to
be a perfectly consistent deformation of a 6D\ SCFT, however, and so such
cases must be addressed in a full analysis of 6D RG flows.

Provided we confine our attention to one particular kind of flow, we can
separately classify the tensor branch, nilpotent, and semi-simple deformations.
The challenge comes when we try to combine these three types of deformations.
If we first perform a tensor branch
flow, we effectively change the endpoint of the configuration. For instance,
we might take the rightmost six curves in the quiver to infinite size,
yielding
\begin{equation}
\lbrack\mathfrak{f}_{4}]\,\,1\,\,\overset{\mathfrak{g}_{2}}{3}%
\,\,\overset{\mathfrak{su}(2)}{2}\,\,2\,\,1\,\,\overset{\mathfrak{e}%
_{8}}{(12)}\,\,\cdots\,\,1\,\,2\,\,\overset{\mathfrak{su}(2)}{2}%
\,\,\overset{\mathfrak{g}_{2}}{3}\,\,1\,\,\overset{\mathfrak{f}_{4}%
}{5}\,\,1\,\,\overset{\mathfrak{g}_{2}}{3}\,\,\overset{\mathfrak{su}%
(2)}{2}\,\,2\,\,1\,\,[\mathfrak{e}_{8}]
\end{equation}
The result is a quiver with endpoint $322...2$. Now, nilpotent deformations on
the left-hand side of the quiver are labeled by nilpotent orbits of the flavor
symmetry $\mathfrak{f}_{4}$ rather than $\mathfrak{e}_{8}$.

A semi-simple deformation of this theory will once again yield a chain of
$\mathfrak{e}_{7}$ gauge algebras. Now, however, the endpoint of the UV theory
is $322...2$, so the IR theory must also have endpoint $322...2$.

The most complicated combination of flows is one in which we first perform a
nilpotent deformation followed by a semi-simple deformation. Note that this is the
reverse of the steps listed below line (\ref{FuriousStyles}). To get a full characterization of
RG flows, however, we need to be able to cover such situations as well.

It is simplest to see how this works in the case of $k$ M5-branes probing a $\mathbb{C}%
^{2}/\mathbb{Z}_{m}$ singularity:
\begin{equation}
\lbrack\mathfrak{su}(m)]\,\,\underset{k - 1}{\underbrace{\overset{\mathfrak{su}%
(m)}{2}\,\,\overset{\mathfrak{su}(m)}{2}\,\,\cdots\,\,\overset{\mathfrak{su}%
(m)}{2}}}\,\,[\mathfrak{su}(m)]
\end{equation}
In this case, nilpotent deformations are labeled by partitions of $m$. Suppose
we first deform the left side of the quiver by some $\mu_{\mathrm{L}}^{(m)}$ (where the superscript indicates that $\mu_{\mathrm{L}}$ is a partition of $m$) and then
perform a semi-simple deformation taking $m\rightarrow m-1$ everywhere in the
quiver. What happens to the left-hand side of the quiver?

Our claim is that this sequence is equivalent to first performing the
semi-simple deformation taking $m\rightarrow m-1$ and \emph{then} deforming the
left-hand side of the quiver by a nilpotent orbit $\mathcal{O}_{\mathrm{L}}^{(m-1)}$ of
$\mathfrak{su}(m-1)$, with $\mathcal{O}_{\mathrm{L}}^{(m-1)}$ given by the maximal
restriction of $\mathcal{O}_{\mathrm{L}}^{(m)}$ to the subalgebra $\mathfrak{su}%
(m-1)\subset\mathfrak{su}(m)$ left unbroken by the semi-simple deformation (we ignore additional $\mathfrak{u}(1)$ factors retained by the semi-simple deformation). In the present case, we arrive at the partition
$\mu_{\mathrm{L}}^{(m-1)}$ by simply deleting the last box in the partition $(\mu
_{\mathrm{L}}^{(m)})^T$. So, for instance, for $m=4$, $\mu_{\mathrm{L}}^{(m)}=[2,2]$, we have
$\mu_{\mathrm{L}}^{(m-1)}=[2,1]$. For $m=7$, $\mu_{\mathrm{L}}^{(m)}=[3,2^{2}]$, we have
$\mu_{\mathrm{L}}^{(m-1)}=[2^3]$. From the perspective of the SCFT quiver, we have
\begin{align}
\overset{\mathfrak{su}(2)}{2}\,\, \underset{[SU(2)]}{\overset{\mathfrak{su}(4)}{2}}  \,\,\overset{\mathfrak{su}(4)}{2}\,\cdots ~&\rightarrow~\underset{[N_f=1]}{\overset{\mathfrak{su}(2)}{2}}\,\, \underset{[N_f=1]}{\overset{\mathfrak{su}(3)}{2}}  \,\,\overset{\mathfrak{su}(3)}{2}\,\cdots \\
\overset{\mathfrak{su}(3)}{2}\,\, \underset{[SU(2)]}{\overset{\mathfrak{su}(6)}{2}}  \,\, \underset{[N_f=1]}{\overset{\mathfrak{su}(7)}{2}} \,\, \overset{\mathfrak{su}(7)}{2} \, \cdots ~&\rightarrow~ {\overset{\mathfrak{su}(3)}{2}}\,\, \underset{[SU(3)]}{\overset{\mathfrak{su}(6)}{2}}  \,\,\overset{\mathfrak{su}(6)}{2}\,\,\overset{\mathfrak{su}(6)}{2}\,\cdots
\end{align}

Similar considerations apply for the case of M5-branes probing a
$\mathbb{C}^{2}/\Gamma_{D_{m}}$ singularity. Nilpotent orbits are labeled by
D-partitions of $2m$, which are partitions of $2m$ subject to the requirement
that any even number in the partition must appear an even number of times.
Performing a nilpotent deformation labeled by some D-partition $\mu_{\mathrm{L}}%
^{(2m)}$ of $2m$ followed by a semi-simple deformation $m\rightarrow m-1$ is
equivalent to first performing the semi-simple deformation followed by a
nilpotent deformation labeled by a D-partition $\mu_{\mathrm{L}}^{(2m-2)}$, which is
obtained by subtracting the last two boxes of $(\mu_{\mathrm{L}}^{(2m)})^T$. The resulting partition $\mu_{\mathrm{L}}^{(2m-2)}$ will still be subject to the
constraint that each even entry appears an even number of times. For
instance, for $2m=10$, $\mu_{\mathrm{L}}^{(2m)}=[3^{2},2^{2}]$, we have $\mu
_{\mathrm{L}}^{(2m-2)}=[2^4]$. For $2m=14$, $\mu_{\mathrm{L}}^{(2m)}=[5,4^2,1]$, we have
$\mu_{\mathrm{L}}^{(2m-2)}=[4^2,3,1]$.

A similar story holds when we have an $\mathfrak{sp}$-type flavor symmetry algebra, which arises after deforming the theory of M5-branes probing a
$\mathbb{C}^{2}/\Gamma_{D_{m}}$ singularity by taking a $-1$ curve to infinite size.
Here, nilpotent orbits of $\mathfrak{sp}(m)$ are labeled by a C-partition of $2m$, which is a partition of $2m$ subject to the constraint that any \emph{odd} number in the partition must appear an even number of times. Once again performing a nilpotent deformation labeled by some C-partition $\mu_{\mathrm{L}}%
^{(2m)}$ of $2m$ followed by a semi-simple deformation $m\rightarrow m-1$ is
equivalent to first performing the semi-simple deformation followed by a
nilpotent deformation labeled by a C-partition $\mu_{\mathrm{L}}^{(2m-2)}$, which is
obtained by subtracting the last two boxes of $(\mu_{\mathrm{L}}^{(2m)})^T$.

Combinations of semi-simple and nilpotent
deformations in theories with exceptional algebras do not have such a simple description in terms of partitions. In addition, we do not presently have a systematic classification of Higgs branch flows involving theories constructed by fusion. Quite possibly, such RG flows
can be understood by determining all ways fusion can be used to generate such a theory. Then, the problem would reduce to the study of
RG flows for each of the decay products which were fused to produce this fusion product. We leave such issues for future work.

\section{Conclusions \label{sec:CONC}}

In this paper we have taken some steps in classifying supersymmetric 6D RG flows,
showing that the vast majority of 6D\ SCFTs can either be viewed as the
products of fission from a small class of progenitor theories. The remaining
small number of theories can be viewed as
fusion products from this fissile material. At the very least, this leads to
an alternative, more algebraic classification scheme which significantly
streamlines the rather rigid structure observed in the classification results
of references \cite{Heckman:2013pva, DelZotto:2014hpa, Heckman:2015bfa}.
More ambitiously, this clear hierarchy
of theories points the way to a corresponding stratification of 6D\ SCFTs.
This is much more data than simply assigning a \textquotedblleft height
function\textquotedblright\ to each 6D\ SCFT (namely the Euler conformal
anomaly). In the remainder of this section we detail some potential areas for
future investigation.

It is remarkable that so many Higgs branch flows can be associated with group
theoretic homomorphisms into some flavor symmetry algebra. But
we have also seen that some complex deformations do not admit such an interpretation, and are instead associated
with semi-simple elements of the flavor symmetry algebra. It would clearly be
worthwhile to systematize the associated flows generated by such deformations,
and in particular, the effects of the most general algebraic deformations
coming from a combination of semi-simple and nilpotent elements of a flavor
symmetry algebra.

One of the interesting results of the present work is that when the number of
maximal gauge algebras on the partial tensor branch is sufficiently large, we
have a uniform characterization of the resulting theories as
fission products of a small class of UV\ progenitor theories. With this in mind,
it would be very interesting to study the holographic RG\ flows associated with fission of
6D SCFTs.

On the other hand, not all theories with a small number of gauge algebras on the tensor branch can be realized as fission products: some of them are ``fusions" of one or more such products, in which a common flavor symmetry is gauged. We have carefully checked that such fusions exist, but we have not found any examples of 6D SCFTs that require more than one such fusion. A systematic classification of fusion products
and their RG flows would be desirable.

The appearance of an interface between data on the left-hand and
right-hand sides of our generalized quivers naturally suggests an interpretation of the degrees
of freedom of 6D SCFTs as edge modes localized on a defect in a higher-dimensional topological theory,
much as in \cite{Heckman:2017uxe}. It would be very interesting to see whether
this observation can be extended to all 6D SCFTs. This would
likely also provide additional insight into topological aspects of 6D SCFTs.

Another natural application of the present work is in the study of
compactifications of 6D\ SCFTs to lower dimensional theories. Indeed, precisely
because so many of these 6D\ SCFTs arise from a simple class of
UV\ progenitors, it is enough to understand the compactification of these
progenitors, and the effects of 6D\ RG\ flows on their lower-dimensional
counterparts. Moreover, since the outlier theories arise from fusion
operations, which are again a canonical operation in lower dimensional
theories, this also paves the way to potentially understanding
compactifications of all 6D\ SCFTs in a rather uniform fashion.

Finally, there is the original ambitious motivation for the present work:\ the
classification of supersymmetric 6D RG flows. Here we have
seen that much of this structure can be boiled down to simple algebraic data.
It would be quite interesting to take even these partial results and
re-interpret them in Calabi-Yau geometry. Indeed, it is likely that much as in
the earlier work on the classification of 6D\ SCFTs, a full classification of
such RG\ flows will involve a tight interplay between algebraic and geometric data.

\newpage

\section*{Acknowledgements}

We thank F.~Apruzzi, D.~Frey, F.~Hassler, N.~Mekareeya, D.R.~Morrison, T.B.~Rochais, and E.~Witten
for helpful discussions. We thank the Banff International Research Station
for hospitality during workshop 18w5190 on the Geometry and Physics of
F-theory. JJH and TR also thank the 2018 Summer Workshop at the Simons Center for Geometry and Physics
for hospitality during the completion of this work.
TR thanks the high energy theory group at the University of
Pennsylvania for hospitality. The work of JJH is supported by NSF CAREER grant
PHY-1756996. The work of TR\ is supported by the Carl P. Feinberg Founders
Circle Membership and by NSF grant PHY-1606531. The work of AT\ is supported
in part by INFN.

\appendix

\bibliographystyle{utphys}
\bibliography{6DRGClass}

\end{document}